\documentclass[%
 reprint,
 amsmath,amssymb,
 aps,
]{revtex4-2}

\usepackage{graphicx}
\usepackage{dcolumn}
\usepackage{bm}
\usepackage{braket}
\usepackage{grffile}

\usepackage{tcolorbox}
\usepackage{mdframed}

\usepackage{amsmath,amssymb}
\usepackage{mathtools}
\usepackage{tikz}
\usetikzlibrary{quantikz}
\usepackage{pgfplots}
\usepackage{subcaption}

\usepackage{cprotect}
\pgfplotsset{compat=1.16}
\usetikzlibrary{positioning}
\usepackage{multirow}
\usepackage{amssymb}
\usepackage{algpseudocode}
\usepackage[boxruled, linesnumbered]{algorithm2e}
\usepackage{siunitx}
\usepackage{hyperref}

\usepackage{listings}
\usepackage{tikz}

\newcommand{\be}{\begin{equation}}
\newcommand{\ee}{\end{equation}}
\newcommand{\bea}{\begin{eqnarray}}
\newcommand{\eea}{\end{eqnarray}}
     
\begin{document}

\preprint{APS/123-QED}

\title{Quantum benefit of the quantum equation of motion for the strongly coupled many-body problem}


\author{Manqoba Q. Hlatshwayo$^{1}$}
\author{John Novak$^{1}$}
\author{Elena Litvinova$^{1, 2, 3}$}
\email{elena.litvinova@wmich.edu}
\affiliation{%
 $^1$ Department of Physics, Western Michigan University, Kalamazoo, MI, 49008, USA\\
 $^2$ Facility for Rare Isotope Beams, Michigan State University, East Lansing, MI, 48824, USA\\
 $^3$ GANIL, CEA/DRF-CNRS/IN2P3, F-14076 Caen, France
}

\date{\today}
\begin{abstract}

We investigate the quantum equation of motion (qEOM), a hybrid quantum-classical algorithm for computing excitation properties of a fermionic many-body system, with a particular emphasis on the strong-coupling regime. The method is designed as a stepping stone towards building more accurate solutions for strongly coupled fermionic systems, such as medium-heavy nuclei, using quantum algorithms to surpass the current barrier in classical computation.  
Approximations of increasing accuracy to the exact solution of the Lipkin-Meshkov-Glick Hamiltonian with $N=8$ particles are studied on digital simulators and IBM quantum devices. Improved accuracy is achieved by applying operators of growing complexity to generate excitations above the correlated ground state, which is determined by the variational quantum eigensolver (VQE).
We demonstrate explicitly that the qEOM exhibits a quantum benefit due to the independence of the number of required quantum measurements from the configuration complexity. Post-processing examination shows that quantum device errors are amplified by increasing configuration complexity and coupling strength. A detailed error analysis is presented, and error mitigation based on zero noise extrapolation is implemented. 
\begin{description}
\item[Keywords] Nuclear Many-Body Problem, Quantum Equation of Motion, Lipkin model
\end{description}
\end{abstract}

\maketitle


\section{\label{sec:intro} Introduction }


During the last several decades, the progress in the nuclear many-body problem has been driven by considerable advancements in its two major building blocks: (i) nucleon-nucleon (NN) interactions and (ii) quantum many-body methods. The former is mainly focused on chiral perturbation theory for bare interactions and density functional theory (DFT) for effective interactions, while the latter develops techniques for modeling the in-medium dynamics of nucleons using NN interactions as an input.
Since the exact many-body solutions for medium-heavy nuclei are beyond the reach of existing computational capabilities, the major goal of theory is to adequately capture the microscopic mechanisms of formation of the emergent collective effects, which considerably redefine the bare forces in strongly correlated media. Such effects are responsible for vibrational and rotational types of motion, superfluidity, collective giant resonances, and more exotic less collective soft modes. An accurate description of emergent collectivity dominating these modes is critical for many nuclear physics applications, ranging from the study of exotic nuclei to astrophysics and the search for new physics.

The success of early phenomenological models that included collective variables established the notion of collective degrees of freedom, such as rotational and vibrational quanta, which can be introduced in the effective Hamiltonians independently of the single-nucleon degrees of freedom \cite{BohrMottelson1969,BohrMottelson1975}. 
Such models, later evolved into more microscopic frameworks, although quite successful, still have not reached the spectroscopic accuracy of even hundreds of keV in the description of excitation spectra and other properties of medium-mass and heavy nuclei. The most advanced approaches belonging to this class encompass various beyond-mean-field (BMF) techniques \cite{BertschBortignonBroglia1983,Soloviev1992,DukelskyRoepkeSchuck1998,Andreozzi2008,GambacurtaGrassoCatara2010,LitvinovaRingTselyaev2008,LitvinovaRingTselyaev2010,Lenske:2019ubp,Tsoneva:2018ven,LitvinovaSchuck2019} implemented on the base of the latest energy density functionals (EDFs) \cite{Bender2003,Decharge1980,Fayans1994,Reinhard2017,MENG2006470,Niksic2011}. Despite the convincing progress on both BMF methods and the EDFs, it remains unclear to what degree the lack of accuracy should be attributed to imperfections of the EDFs, inherent deficiencies of the BMF models, or unavoidable limitations of present computational capabilities.

Some hope to resolve these issues laid with 'ab initio' approaches, which were supposed to use, as their only input, the NN interaction in the vacuum. However, such approaches (dominated by chiral perturbation theory combined with standard many-body methods \cite{Hergert2016,Hagen2014,Soma2013,Cipollone2013}) leave it unclear to what extent the emergent collectivity, crucial for medium-heavy nuclei, can be addressed. Presently available calculations of this kind require re-adjustments of the NN-interaction to the properties of finite nuclei \cite{Ekstrom2015,Miyagi2020}, thereby partly absorbing the collective effects in the parameters. 

The complexity of the nuclear many-body problem is a serious obstacle on the way to a spectroscopically accurate theory, which considerably impedes its utility in applications where highly accurate computation for medium-mass and heavy nuclei is crucial. Many-body response theory is the best tool to quantify the excitation spectra of such nuclei, and the best-quality nuclear response calculations beyond the simplistic (quasiparticle) random phase approximation ((Q)RPA) only include configuration complexity up to (correlated) two-particle-two-hole ($2p2h$) \cite{LitvinovaRingTselyaev2008,LitvinovaRingTselyaev2010,Gambacurta2011,Gambacurta2015,NiuColoVigezziEtAl2014,NiuNiuColoEtAl2015,RobinLitvinova2018,Tselyaev2018,Robin2019}, and in rare cases $3p3h$ \cite{Ponomarev1999,LoIudice2012,Savran2011,Tsoneva:2018ven,Lenske:2019ubp}, due to limitations in current computational capabilities. These implementations are mostly based on effective NN-interactions, either schematic or derived from density functional theories (DFTs), although calculations employing bare interactions have also become available \cite{PapakonstantinouRoth2009,Bacca:2013dma,Knapp:2014xja,Bacca2014,Knapp:2015wpt,DeGregorio2016a,Raimondi2018}. Both types of approaches to the nuclear response still do not demonstrate consistent performance on spectroscopic accuracy, and the absence of a clear link to the exact equations of motion (EOMs) for the fermionic correlation functions somewhat obscures their assessment. A recent effort to surmount these shortcomings has been the advancement of relativistic nuclear field theory in a fully self-consistent approach to the nuclear response based on the exact EOMs for fermionic propagators and including correlated $3p3h$ configurations in a large model space \cite{LitvinovaSchuck2019,Litvinova2023a}. These calculations emphasize the importance of taking into account high-complexity configurations in the parameter-free formalism and pave the way to a spectroscopically accurate and yet computationally feasible theory of nuclear spectral properties that satisfy the standards of modern applications.

However, even with the advent of exascale computing, the impediment of exponential growth of the Hilbert space with configurations of growing complexity remains a serious hurdle in classical numerical approaches. Therefore, quantum algorithms have become an attractive alternative for practitioners, especially because the nuclear many-body problem represents a category for which quantum computers may outperform traditional ones. 
While a fully coherent universal quantum computer remains unrealized, there exist algorithms suitable for the presently available noisy intermediate-scale quantum (NISQ) devices. In the nuclear physics domain, during the last few years, several theoretical groups have reported applications of such algorithms to nuclear systems and relevant model Hamiltonians, addressing both their static and dynamic properties. The former, which are based on the variational quantum eigensolver, include computing the binding energy of light nuclei \cite{Dumitrescu2018,Lu2019a} and simulation of lattice models \cite{Kokail2019}. The latter includes a quantum algorithm for linear response theory \cite{Roggero2019}, the time evolution of a nuclear many-body system \cite{guzman2021calculation}, and simulation of non-Abelian gauge theories with optical lattices \cite{Tagliacozzo2013}. Other efforts in the field address efficient state preparation schemes \cite{Roggero2020,Lacroix2020,Guzman2022} and analysis of nuclear structure using entanglement \cite{Robin2021,robin2023quantum}.

The quantum EOM (qEOM) algorithm, first implemented for quantum chemistry calculations \cite{Ollitrault2020} as a quantum extension of the VQE method ~\cite{Tilly2022}, has attracted our attention as it is best aligned with the strategy of building growing-complexity solutions to the fermionic many-body problem with controlled uncertainties, in the spirit of Refs. \cite{LitvinovaSchuck2019,Litvinova2023a}. 
Furthermore, the qEOM algorithm is found to be more resilient to noise compared with other currently available methods, especially in its recently updated self-consistent version \cite{Asthana2023}, and it bypasses using a quantum computer for the noise-sensitive step of solving the eigenvalue equation, leaving it to a classical computer. 
While this method was originally only applied to molecular calculations, i.e., for weakly-coupled systems, in Ref. \cite{Hlatshwayo2022}, we explored its performance in strongly-coupled regimes of a prototype Lipkin-Meshkov-Glick (LMG) Hamiltonian, also known as Lipkin model. It was found that qEOM shows a very reasonable performance and has a large potential for improvement.

In this work, we expand the study of Ref. \cite{Hlatshwayo2022} in various aspects. 
We note that the accuracy of the qEOM largely depends on the configuration complexity $\alpha_{m}$ of the excitation operator $O^\dagger_{n}(\alpha_m)$ where $n$ denotes the $n^{\text{th}}$ excited state. In general, an $N$-body problem requires complexity $\alpha_m = N$ for the exact solution, whereas the most advanced classical computation of medium-mass and heavy nuclei ($N \approx 100$) hardly reaches the complexity of $\alpha_m = 3$. 
We show that the qEOM exhibits a quantum benefit when building a hierarchy of approximations ordered by $\alpha$ due to the independence of the number of required quantum measurements on this parameter. The number of measurements is found to scale with the number of particles, thus given this number one can increase the configuration complexity and achieve better accuracy of the qEOM without performing more measurements on the quantum computer. 
Using the efficient encoding scheme introduced in Ref. \cite{Hlatshwayo2022}, we show that the number of required quantum measurements scales at most quadratically with the number of particles for the Lipkin model, implying that the method is feasible for large systems. Furthermore, we show that for a given particle number, one can increase the configuration complexity and achieve better accuracy of the qEOM without performing more measurements on the quantum computer. 

To demonstrate this capability, we simulate the energy spectrum 
with $\alpha_m = \{1, 2, 3 \}$ using IBM quantum computers for the LMG Hamiltonian with $N=8$ particles and running coupling strength. The qEOM results for $\alpha_m=3$, after performing error mitigation, show a reasonable agreement with the exact solution even in the domain of strong coupling. The sensitivity of the algorithm to the decoherence of quantum hardware and sampling noise is discussed.


\section{Equation of motion methods} 

In this Section, we bridge two EOM methods, which are essentially equivalent but usually employed in different contexts. The first EOM method is the backbone of the response theory dealing with EOMs for fermionic propagators \cite{AdachiSchuck1989,DukelskyRoepkeSchuck1998}, the major subject of study in nuclear and particle physics. The second method is the EOM of Rowe \cite{rowe1968equations}, which directly targets the excited states of the many-body quantum system and sets a common background for nuclear structure and quantum chemistry. The latter EOM serves as a foundation for the quantum EOM algorithm \cite{Ollitrault2020} utilizing the efficient computation of the many-body ground states by VQE. Here, we extend it to the computation of the transition matrix elements, thereby establishing a link with the response theory and accentuating the universality of the approach and its applicability across the subfields of quantum many-body physics.

\subsection{Response theory}\label{sec:response} 

The strength (spectral) function, which determines the excitations of a fermionic system in response to a weak oscillating external field $F^{\dagger}$ is defined by Fermi's golden rule:
\bea
S(\omega) = \sum\limits_{\nu>0} \Bigl[ |\langle \nu|F^{\dagger}|0\rangle |^2\delta(\omega-\omega_{\nu}) - |\langle \nu|F|0\rangle |^2\delta(\omega+\omega_{\nu})
\Bigr], \nonumber\\
\label{SF}
\eea
with the summation over all excited states $|\nu\rangle$. The transition matrix element $\langle \nu|F^{\dagger}|0\rangle$, for the one-body external field operator in the second-quantized form, reads:
\be
\langle \nu|F^{\dagger}|0\rangle = \sum\limits_{12}\langle \nu|F_{12}^{\ast}\psi^{\dagger}_2\psi_1|0\rangle = \sum\limits_{12}F_{12}^{\ast}\rho_{21}^{\nu\ast}.
\label{Frho}
\ee
The matrix elements in Eq. (\ref{Frho}) are the transition densities
\be
\rho^{\nu}_{12} = \langle 0|\psi^{\dagger}_2\psi_1|\nu \rangle , 
\label{trden}
\ee
which can be interpreted as the weights of the pure particle-hole configurations $\psi^{\dagger}_2\psi_1$ in the single-particle basis $\{1\}$, built on top of the ground state $|0\rangle$, in the excited states $|\nu\rangle$. 
Approximating the delta-functions in Eq. (\ref{SF}) by the Lorentz distribution
\be
\delta(\omega-\omega_{\nu}) = \frac{1}{\pi}\lim\limits_{\Delta \to 0}\frac{\Delta}{(\omega - \omega_{\nu})^2 + \Delta^2},
\ee
one obtains
\be
S(\omega) 
= -\frac{1}{\pi}\lim\limits_{\Delta \to 0} {\Im} \Pi(\omega+ i\Delta),
\label{SFDelta} 
\ee
where $\Pi(\omega)$ is the polarizability of the many-body system:
\be
\Pi(\omega) 
=  \sum\limits_{\nu} \Bigl[ \frac{B_{\nu}}{\omega - \omega_{\nu}}
- \frac{{\bar B}_{\nu}}{\omega + \omega_{\nu} }
\Bigr]
\label{Polar}
\ee
related to the transition probabilities $B_{\nu}$ and ${\bar B}_{\nu}$ of absorption and emission, respectively:
\be
B_{\nu} = |\langle \nu|F^{\dagger}|0\rangle |^2\ \ \ \ \ \ \ \ 
{\bar B}_{\nu} = |\langle \nu|F|0\rangle |^2
\label{Prob}.
\ee
Thus, the strength function associated with the given external field operator $F$ can be expressed as
\be
S_F(\omega) = -\frac{1}{\pi}\lim_{\Delta\to 0}\Im\sum\limits_{121'2'}F_{12}R_{12,1'2'}(\omega+i\Delta)F^{\ast}_{1'2'},
\label{SFF}
\ee
where the response function $R_{12,1'2'}(\omega)$ is figuring in its spectral representation
\be
R_{12,1'2'}(\omega) = \sum\limits_{\nu>0}\Bigl[ \frac{\rho^{\nu}_{21}\rho^{\nu\ast}_{2'1'}}{\omega - \omega_{\nu} + i\delta} -  \frac{\rho^{\nu\ast}_{12}\rho^{\nu}_{1'2'}}{\omega + \omega_{\nu} - i\delta}\Bigr],
\label{respspec}
\ee
with the poles at the energies $\omega_{\nu} = E_{\nu} - E_0$ of the excited states relative to the ground state energy and $\delta \to +0$. 

The response function can be thus defined in terms of the time-dependent fermionic field operators as the two-time correlation function
\be
R(12,1'2') \equiv R_{12,1'2'}(t-t') =  -i\langle T\psi^{\dagger}(1)\psi(2)\psi^{\dagger}(2')\psi(1')\rangle,
\label{phresp}
\ee
where $\braket{.}$ is a shorthand notation for the expectation value in the ground state while
\bea
\psi(1) \equiv \psi_1(t_1) \equiv {e}^{iHt_1}\psi_1 {e}^{-iHt_1} \nonumber \\ 
\psi^{\dagger}(1) \equiv \psi^{\dagger}_1(t_1) \equiv {e}^{iHt_1}\psi^{\dagger}_1 {e}^{-iHt_1}.
\label{t-fields}
\eea 
In Eqs. (\ref{phresp},\ref{t-fields})
it is implied that the number arguments in the brackets include the time variables, $t_1 = t_2 = t$ and $t_{1'} = t_{2'} = t'$. The operator $H$ in Eq. (\ref{t-fields})
is the many-body fermionic Hamiltonian
\be
H = H^{(1)} + V^{(2)}
\label{Hamiltonian}
\ee
confined here by the two-body interaction $V^{(2)}$.
In the one-body term $H^{(1)}$
\be
H^{(1)} = \sum_{12} t_{12} \psi^{\dag}_1\psi_2 + \sum_{12}v^{(MF)}_{12}\psi^{\dag}_1\psi_2 \equiv \sum_{12}h_{12}\psi^{\dag}_1\psi_2
\label{Hamiltonian1}
\ee
the matrix elements $h_{12}$ combine the kinetic and the mean-field $v^{(MF)}$ parts of the interaction. The two-body sector is defined by the operator $V^{(2)}$
\be
V^{(2)} = \frac{1}{4}\sum\limits_{1234}{\bar v}_{1234}{\psi^{\dagger}}_1{\psi^{\dagger}}_2\psi_4\psi_3,
\label{Hamiltonian2}
\ee
via ${\bar v}_{1234} = v_{1234} - v_{1243}$, the antisymmetrized matrix element of the interaction of two fermions in the vacuum, also called the bare interaction.
Three-body forces are neglected in this work but can be included straightforwardly.  

The EOM for the response function (\ref{phresp}) is generated by differentiation of (\ref{phresp}) with respect to the time arguments. Taking the derivative  with respect to $t$, one obtains:
\bea
(i\partial_t + \varepsilon_{12})R_{12,1'2'}(t-t')
= \delta(t-t'){\cal N}_{121'2'} 
\nonumber \\
+ i\langle T[V,{\psi^{\dagger}}_1\psi_2](t)({\psi^{\dagger}}_{2'}\psi_{1'})(t')\rangle ,
\label{dtG2b}                           
\eea
where ${\cal N}_{121'2'}$ is the norm kernel:
\bea
{\cal N}_{121'2'} = \langle[\psi^{\dagger}_{1}\psi_{2},\psi^{\dagger}_{2'}\psi_{1'}]\rangle =  \delta_{22'}\langle \psi^{\dagger}_{1}\psi_{1'} \rangle - 
\delta_{11'}\langle \psi^{\dagger}_{2'}\psi_{2} \rangle. 
\nonumber \\
\label{norm}
\eea
In Eq. (\ref{dtG2b}) and in the following $\varepsilon_{12} = \varepsilon_{1} - \varepsilon_{2}$, while $\varepsilon_{1}$ and $ \varepsilon_{2}$ are the eigenvalues of the
one-body part of the Hamiltonian. The basis single-particle states are, therefore, the eigenstates of $H^{(1)}$ and $h_{12} =  \delta_{12}\varepsilon_1$. 
The norm simplifies to the form: ${\cal N}_{121'2'} = \delta_{11'}\delta_{22'}(n_1 - n_2) \equiv \delta_{11'}\delta_{22'}{\cal N}_{12}$, where
$n_1 = \langle{\psi^{\dagger}}_1\psi_1\rangle$ is associated with the occupancy of the fermionic state $|1\rangle$. 

The differentiation of the last term on the right-hand side of Eq. (\ref{dtG2b}) with respect to $t'$ generates the second EOM:
\bea
i\langle T[V,{\psi^{\dagger}}_1\psi_2](t)({\psi^{\dagger}}_{2'}\psi_{1'})(t')\rangle(-i\overleftarrow{\partial_{t'}} - \varepsilon_{2'1'}) =
\nonumber \\
= -\delta(t-t')
\langle [[V,{\psi^{\dagger}}_1\psi_2],{\psi^{\dagger}}_{2'}\psi_{1'}]\rangle + \nonumber \\ + i\langle T[V,{\psi^{\dagger}}_1\psi_2](t)[V,{\psi^{\dagger}}_{2'}\psi_{1'}](t')\rangle.
\label{dtG2c}
\eea
The combination of Eqs. (\ref{dtG2c}) and (\ref{dtG2b}), after a Fourier transformation to the energy domain, yields
\bea
R_{12,1'2'}(\omega) = R^{(0)}_{12,1'2'}(\omega) \nonumber\\
+ \sum\limits_{343'4'} R^{(0)}_{12,34}(\omega){\cal T}_{34,3'4'}(\omega)R^{(0)}_{3'4',1'2'}(\omega),
\label{2bgfb}
\eea
with the free (uncorrelated) particle-hole response $R^{(0)}(\omega)$
\be
R^{(0)}_{12,1'2'}(\omega) = \frac{{\cal N}_{121'2'}}{\omega - \varepsilon_{21}} = \delta_{11'}\delta_{22'}\frac{n_1 - n_2}{\omega - \varepsilon_{21}} 
\label{resp0}
\ee
and the $T$-matrix ${\cal T}(\omega)$, the Fourier image of ${\cal T}(t-t')$ which splits into the instantaneous $T^{(0)}$ and the time-dependent ${\cal T}^{(r)}$ parts:
\bea
{\cal T}_{12,1'2'}(t-t') &=& {\tilde{\cal N}}_{121'2'}^{-1}\Bigl({\cal T}^{(0)}_{12,1'2'}\delta(t-t') + {\cal T}^{(r)}_{12,1'2'} (t-t')\Bigr), \nonumber \\ 
{\tilde{\cal N}}_{121'2'} &=& {\cal N}_{12}{\cal N}_{1'2'}, \\
{\cal T}^{(0)}_{12,1'2'} &=& -\langle [[V,{\psi^{\dagger}}_1\psi_2],{\psi^{\dagger}}_{2'}\psi_{1'}]\rangle, \nonumber \\
{\cal T}^{(r)}_{12,1'2'}(t-t') &=&  i\langle T[V,{\psi^{\dagger}}_1\psi_2](t)[V,{\psi^{\dagger}}_{2'}\psi_{1'}](t')\rangle .\nonumber\\
\label{Ft} 
\eea
By introducing the irreducible interaction kernel $K(t-t')$, where irreducibility is implied with respect to $R^{(0)}_{12,1'2'}$,
\bea
K(t-t') = {\tilde{\cal N}}^{-1}\Bigl(K^{(0)}\delta(t-t') + K^{(r)}(t-t')\Bigr), \nonumber \\ 
K^{(0)} = {\cal T}^{(0)}, \ \ \ \ \ \  K^{(r)}(t-t') = {\cal T}^{(r)irr}(t-t'),\nonumber \\
\label{Womega}
\eea 
Eq. (\ref{2bgfb}) transforms to a formally closed equation for $R(\omega)$, similar to the Dyson equation for the one-fermion 
propagator, which is known as the Bethe-Salpeter-Dyson equation (BSDE):
\be
R(\omega) = R^{(0)}(\omega) + R^{(0)}(\omega)K(\omega)R(\omega),
\label{BSDE}
\ee
where
\be
{\cal T}(\omega) = K(\omega) + K(\omega)R^{(0)}(\omega){\cal T}(\omega). 
\label{Kkernel}
\ee
Eqs. (\ref{resp0}-\ref{Kkernel}) express the response theory, which can be applied for calculations of the strength distribution (\ref{SF}) for a given $F^{\dagger}$. The excitation spectrum of the system is, thereby, in principle, completely determined by the external field and the bare fermionic interaction via the commutation relations (\ref{norm},\ref{Ft}), which promote all the in-medium physics.
Because of the presence of higher-rank correlation functions in the 
dynamical, or time-dependent, interaction kernel $K^{(r)}(t-t')$ which give a feedback on its static counterpart,
in practice, certain approximations are made to obtain solutions of Eq. (\ref{BSDE}) \cite{LitvinovaSchuck2019,Schuck2021}. 

At the exact poles of the response function $\omega \to \omega_{\nu}$ BSDE can be reformulated into a matrix equation for the transition densities
\be
\rho^{\nu}_{21} = \sum\limits_{341'2'}R^{(0)}_{12,34}(\omega_{\nu})K_{34,1'2'}(\omega_{\nu})\rho^{\nu}_{2'1'}.
\ee
Under certain assumptions about the correlation content of the ground and excited states, this relationship can be further recast into the form of the generalized eigenvalue equation, which appears in the Rowe's EOM method discussed in the next subsection. 


\subsection{Quantum EOM algorithm and its quantum benefit}\label{sec:qeom_benefit}

The EOM developed by Rowe \cite{rowe1968equations} is another framework for computing excitation properties of quantum many-body systems. The goal is
to find the excitation spectrum of a quantum system obeying the Schr\"{o}dinger equation
\begin{equation}\label{Eqn:schrodinger}
    \hat{H} \ket{n} = E_n \ket{n},
\end{equation} 
where $\hat{H}$ is the Hamiltonian operator, $E_n$ is the energy of the $n^{th}$ excited state, and $\ket{n}$ is the $n^{th}$ excited state.
Following tradition, in this subsection, we will denote operators explicitly with the '~$\hat{}$~' symbol.
The excited states $\ket{n}$ are generated by the excitation operator $\hat{O}^\dagger_n$ which is defined by its action on the ground state $\ket{0}$:
\begin{equation}
  \ket{n} = \hat{O}^\dagger_n \ket{0}  
  \label{ExOp}
\end{equation}
with the vacuum annihilation condition (VAC)
\be
\hat{O}_n \ket{0} = 0.
\label{vacuum}
\ee
The energies above the ground-state energy ($E_{n0} = E_n - E_0 \equiv \omega_n $) are given by Ref. \cite{RingSchuck1980} as
\begin{equation}\label{Eqn:EOM}
      E_{n0} = \frac{ \braket{\left[ \hat{O}_n, \left[ \hat{H}, \hat{O}^\dagger_n \right]  \right]}  }{ \braket{\left[ \hat{O}_n, \hat{O}^\dagger_n \right]} }.
\end{equation}
Therefore, the task of solving Eq. (\ref{Eqn:schrodinger}) has been reformulated into calculating the energy spectrum Eq. (\ref{Eqn:EOM}) and the associated wave functions of Eq. (\ref{ExOp}) using knowledge of the excitation operator $\hat{O}^\dagger_n$ and the many-body ground-state $\ket{0}$. If $\hat{O}^\dagger_n$ or $\ket{0}$ is not known exactly, some reasonable techniques to approximate them can be used to initiate the computation cycle. 

The excitation operator can be written in its most general form as an expansion over products of creation and annihilation field operators 
    \begin{equation}\label{Eqn:excitation_op}
         \hat{O}^\dagger_n (\alpha_m)= \sum_{\alpha = 1}^{\alpha_m} \sum_{\mu_{\alpha}}    \left[ X^{\alpha}_{\mu_{\alpha}} (n) \hat{K}^{\alpha}_{\mu_{\alpha}} - Y^{\alpha}_{\mu_{\alpha}} (n) \left(\hat{K}^{\alpha}_{\mu_{\alpha}}\right)^\dagger \right],   
    \end{equation}
where 
\be
\hat{K}^{1}_{\mu_{1}} = {\hat\psi}^{\dagger}_{p}{\hat\psi}_{h},  \ \ \ \ \ \ \ \hat{K}^{2}_{\mu_{2}} = {\hat\psi}^{\dagger}_{p}{\hat\psi}^{\dagger}_{p'}{\hat\psi}_{h'}{\hat\psi}_{h}, \ \ \ ...,
\ee
the indices $p, h, p', h'$ stand for the particle and hole states (above and below the Fermi surface, respectively), $\alpha$ is the degree of configuration complexity equal to the number of particle-hole pairs, and $\mu_{\alpha}$ is the collective index combining the single-particle states involved in the given configuration. In principle, with $\alpha_{m} = N$, the excitation operator in Eq. (\ref{Eqn:excitation_op}) can generate exact solutions for the $N$-body system assuming that $\ket{0}$ is exact. Thus, for $\alpha_m < N$, there is a hierarchy of approximations that can be related to the Bogoliubov-Born-Green-Kirkwood-Yvon (BBGKY) hierarchy \cite{Bogoliubov1947}. 

In general, many-body systems dominated by strong coupling require higher configuration complexity $\alpha_m$ of $\hat{O}^\dagger_n$ to achieve certain accuracy than the weakly-interacting systems. At the same time, for non-perturbative theories that do not rely on expansions in small parameters, building a hierarchy of approximations with varying $\alpha_m$ may serve as uncertainty quantification, which is an important theory ingredient \cite{piarulli2023}. 
Realistic calculations for medium-heavy nuclear systems \cite{LitvinovaSchuck2019} indicate that the quality of description grows relatively quickly with $\alpha_m$, and the spectral results saturate with this parameter.
Establishing a quantitative link between accuracy and configuration complexity would be of great value because the required accuracy of a particular application (or comparison to experimental data) would set the upper limit on the configuration complexity of the corresponding EOM calculation. This strategy can be implemented with an adaptive algorithm as an alternative to the brute force diagonalization of $\hat{H}$, variants of which are commonly called shell models  \cite{zelevinsky2017physics}. 
Furthermore, as previously noted, the EOM method can be conveniently converted into a quantum algorithm qEOM and used in combination with the VQE method, which efficiently computes the many-body ground state $\ket{0}$ on a quantum computer \cite{Ollitrault2020,Asthana2023}. 

With a good approximation to $\ket{0}$ and a fixed $\hat{O}^\dagger_n$, one can solve Eq. (\ref{Eqn:EOM}) by minimization, i.e., setting the variation $\delta (E_{n0}) = 0$ in the parameter space spanned by the coefficients of Eq. (\ref{Eqn:excitation_op}). This procedure leads to the generalized eigenvalue equation (GEE), which, in the block-matrix form, reads:
\begin{equation}\label{Eqn:GEE}
	\begin{pmatrix} \mathcal{A} &  \mathcal{B} \\ \mathcal{B}^{*} & \mathcal{A}^{*} \end{pmatrix}
	\begin{pmatrix} X_{n} \\ Y_{n} \end{pmatrix}
	=
	E_{n0}
	\begin{pmatrix} \mathcal{C} & \mathcal{D} \\ -\mathcal{D}^{*} & -\mathcal{C}^{*} \end{pmatrix}
	\begin{pmatrix} X_{n} \\ Y_{n} \end{pmatrix}.
\end{equation}
The matrix elements are found by taking expectation values of the following commutators in the ground-state $\ket{0}$:
\begin{eqnarray}
	 \mathcal{A}_{\mu_\alpha \nu_\beta} &=& \braket{ \left[ \left(\hat{K}^{\alpha}_{\mu_{\alpha}}\right)^\dagger, \left[ \hat{H}, \hat{K}^{\beta}_{\nu_{\beta}} \right]  \right] }, 	\qquad  \qquad \; \; 
	 \\
	 \mathcal{C}_{\mu_\alpha \nu_\beta} &=& \braket{ \left[ \left(\hat{K}^{\alpha}_{\mu_{\alpha}}\right)^\dagger, \hat{K}^{\beta}_{\nu_{\beta}}  \right]}, 	\\ 
	 \mathcal{B}_{\mu_\alpha \nu_\beta} &=& - \braket{ \left[ \left(\hat{K}^{\alpha}_{\mu_{\alpha}}\right)^\dagger, \left[ \hat{H}, \left(\hat{K}^{\beta}_{\nu_{\beta}} \right)^\dagger\right]  \right] }, \qquad 
	 \\
	 \mathcal{D}_{\mu_\alpha \nu_\beta} &=& - \braket{ \left[ \left(\hat{K}^{\alpha}_{\mu_{\alpha}}\right)^\dagger,  \left(\hat{K}^{\beta}_{\nu_{\beta}} \right)^\dagger  \right] }. 
\end{eqnarray}
The GEE (\ref{Eqn:GEE}) may be left for post-processing on a classical computer, for which methods of increasing efficiency have been developed \cite{Nakatsukasa2007,Bjelcic2020,Litvinova2022a}. 

The major advantage of the qEOM algorithm resides in the efficient evaluation of the required expectation values for the GEE \cite{Ollitrault2020}.  Another aspect of the qEOM's computational efficiency manifests in simulations of strongly coupled many-body systems, which require higher configuration complexity to obtain accurate results. Specifically, increasing the configuration complexity for a fixed system size is possible without performing additional measurements on a quantum computer. To illustrate this statement, note that the Hamiltonian of a fermionic system can be expressed in terms of a product of Pauli gates $P_k \in \{ \mathbb{I}, X, Y, Z\}$ as follows
 \begin{equation}
	     {H} = \sum_{i} h_{i} \; \{P_0 \otimes P_1 \otimes \ldots P_{n_q}\}^i,
\label{PauliH}      
\end{equation}
where $n_q$ is the number of qubits. Therefore, the basis excitation operators and the matrix elements of $\mathcal{A}$ are given by
\begin{equation}
	     {K} = \sum_{i} \kappa_{i} \; \{P_0 \otimes P_1 \otimes \ldots P_{n_q}\}^i,
\end{equation}
 \begin{equation}
		 \mathcal{A} = \sum_{i} a_{i} \braket{ \{P_0 \otimes P_1 \otimes \ldots P_{n_q}\}^i},
\label{PauliA}
	 \end{equation}
and analogous decompositions can be employed for the other matrices $\mathcal{B}$, $\mathcal{C}$, and $\mathcal{D}$.
The quantum benefit is that the number of qubits $n_q$ is fixed by the particle number $N$; thus, the number of measurements on a quantum computer $\braket{ P_0 \otimes P_1 \otimes \ldots P_q}$ is fixed.  Increasing the configuration complexity $ \alpha$ increases the number of terms in the sums (\ref{PauliH}-\ref{PauliA}) but does not affect the number of measurements. 
This implies an efficient scheme for building a hierarchy of approximations
ordered by the degree of configuration complexity that is needed for realistic nuclear structure computation within the EOM framework.

The GEE yields the complete set of excitation energies $E_{n0}$ and amplitudes $X^{\alpha}_{\mu_{\alpha}}, Y^{\alpha}_{\mu_{\alpha}}$ which can be used to compute the strength function (\ref{SF}) associated with a given external field $\hat F$. The relevant matrix element (e.g., the complex conjugate of Eq. (\ref{Frho})) can be expressed via an expectation value of the commutator of the external field and excitation operators
\be
\bra{0}\hat{F}\ket{n} = \bra{0}[\hat{F},\hat{Q}^{\dagger}_n(\alpha_m)]\ket{0} 
\label{ME}
\ee
by employing the VAC (\ref{vacuum}). As in Eq. (\ref{Eqn:EOM}), the commutator form is used to reduce the rank of operators acting on the correlated ground state. 
The general second-quantized form of a one-body operator
\be
{\hat F} = \sum\limits_{ij}F_{ij}{\hat \psi}^{\dagger}_i{\hat \psi}_j, 
\label{Fhat}
\ee
formally contains all types of contributions: $F_{ph}$, $F_{hp}$, $F_{pp}$, and $F_{hh}$ (where the Latin subscripts in Eq. (\ref{Fhat}) mark the same single-particle basis denoted by number indices in Section \ref{sec:response}). The commutator in (\ref{ME}) thus expands as
\bea
[{\hat F},{\hat Q}^{\dagger}_n(\alpha_m)] &=& 
\sum\limits_{ij}F_{ij}\sum_{\alpha = 1}^{\alpha_m} \sum_{\mu_{\alpha}}    \Bigl( X^{\alpha}_{\mu_{\alpha}} (n) [{\hat \psi}^{\dagger}_i{\hat \psi}_j,\hat{K}^{\alpha}_{\mu_{\alpha}}]  \nonumber \\
&-& Y^{\alpha}_{\mu_{\alpha}} (n) [{\hat \psi}^{\dagger}_i{\hat \psi}_j,\hat{K}^{\alpha\dagger}_{\mu_{\alpha}}] 
\Bigr),   
\eea
with, for $\alpha=1$ and $\alpha=2$:
\be
[{\hat \psi}^{\dagger}_i{\hat \psi}_j,\hat{K}^{1}_{\mu_{1}}] \equiv
[{\hat \psi}^{\dagger}_i{\hat \psi}_j,{\hat\psi}^{\dagger}_{p}{\hat\psi}_{h}] =
\delta_{pj}{\hat\psi}^{\dagger}_{i}{\hat\psi}_{h} - \delta_{hi}{\hat\psi}^{\dagger}_{p}{\hat\psi}_{j}, \nonumber
\ee
\bea
[{\hat\psi}^{\dagger}_i{\hat \psi}_j,\hat{K}^{2}_{\mu_{2}}] \equiv
[{\hat \psi}^{\dagger}_i{\hat \psi}_j,{\hat\psi}^{\dagger}_{p}{\hat\psi}^{\dagger}_{p'}{\hat\psi}_{h'}{\hat\psi}_{h}]  \nonumber \\
= -\delta_{ih}{\hat\psi}^{\dagger}_{p}{\hat\psi}^{\dagger}_{p'}{\hat\psi}_{h'}{\hat\psi}_{j} +
\delta_{ih'}{\hat\psi}^{\dagger}_{p}{\hat\psi}^{\dagger}_{p'}{\hat\psi}_{h}{\hat\psi}_{j} \nonumber \\
-\delta_{jp'}{\hat\psi}^{\dagger}_{i}{\hat\psi}^{\dagger}_{p}{\hat\psi}_{h'}{\hat\psi}_{h} +
\delta_{jp}{\hat\psi}^{\dagger}_{i}{\hat\psi}^{\dagger}_{p'}{\hat\psi}_{h'}{\hat\psi}_{h},
\eea
the analogous expressions for $\alpha \geq 3$ 
and respective counterparts with $p\leftrightarrow h$, $p'\leftrightarrow h', ... $.
The ground-state expectation values of these commutators depend on the correlation content of the model ground state.
For instance, with the uncorrelated ground state of the Hartree or Hartree-Fock (HF) type $\ket{0} = \ket{\text{HF}}$, which is defined as the particle vacuum ${\hat\psi}_p\ket{\text{HF}} = 0$, one obtains
\be
\bra{0}\hat{F}\ket{n} = \sum\limits_{ph} \Bigl( F_{hp}X^1_{ph}(n) + F_{ph}Y^1_{ph}(n)\Bigr).
\label{TrME0}
\ee
This means that in the HF approximation to the ground state $\rho^n_{ph} = X^1_{ph}(n)$, 
$\rho^n_{hp} = Y^1_{ph}(n)$, and only $\alpha = 1$ amplitudes contribute to the transition probabilities
even in the presence of higher-complexity configurations in the operator ${\hat Q}^{\dagger}_n(\alpha_m)$. Eq. (\ref{TrME0}) is most commonly used in response theory and neglects the ground state correlations. 

The simplest and often most relevant is the external field operator of one-body character with non-vanishing $ph$ and $hp$ components:
\be
\hat{F}_0 = \sum\limits_{ph}(F_{ph}{\hat\psi}^{\dagger}_{p}{\hat\psi}_{h} + F_{hp}
{\hat\psi}^{\dagger}_{h}{\hat\psi}_{p}). 
\ee
In this case, the transition amplitudes read:
\bea
\bra{0}\hat{F}_0\ket{n} &=& 
\sum\limits_{ph}\Bigl( F_{ph}\sum_{\alpha = 1}^{\alpha_m} \sum_{\mu_{\alpha}}    \bigl( X^{\alpha}_{\mu_{\alpha}} (n) \langle[\hat{K}^{1}_{ph},\hat{K}^{\alpha}_{\mu_{\alpha}}]\rangle  \nonumber \\
&-& Y^{\alpha}_{\mu_{\alpha}} (n) \langle[\hat{K}^{1}_{ph},\hat{K}^{\alpha\dagger}_{\mu_{\alpha}}]\rangle  
\bigr) + \bigl( p\leftrightarrow h \bigr) \Bigr) ,
\label{TrME}
\eea
where the expectation values on the right-hand side are taken in the formally exact ground state. In the VQE+qEOM approach, we deal with correlated ground states of unspecified correlation content. However, the expectation values entering Eq. (\ref{TrME}) are already contained in the set of the Pauli strings measured for constructing the GEE matrix since 
\be
\langle[\hat{K}^{1}_{ph},\hat{K}^{\alpha}_{\mu_{\alpha}}]\rangle  = {\cal D}^{\ast}_{ph,\mu_{\alpha}}
\ \ \ \ \ \
\langle[\hat{K}^{1}_{ph},\hat{K}^{\alpha\dagger}_{\mu_{\alpha}}]\rangle  = - {\cal C}^{\ast}_{ph,\mu_{\alpha}},   
\ee
so that the quantum advantage extends to the computation of the transition amplitudes and strength functions (\ref{SF}).

Two-body and higher-rank terms can also be present in the external field operators. Of particular interest are two-body currents, which are expressed by two-body operators and can, therefore, non-trivially couple to the correlated ground state. Strength function studies on a quantum computer for various types of external fields will be considered in a separate publication.

\section{qEOM application to Lipkin model} 
\subsection{\label{sec:lipkin} Lipkin Model}
The LMG model is a test-bed for approximate techniques of solving fermionic quantum many-body problem~\cite{lipkin1965validity, meshkov1965validity, glick1965validity, Vidal2007, Vidal2008}.
It describes a system of $N$ interacting fermions constrained to two $N$-fold degenerate energy levels with $E = \pm\epsilon/2$. The particles interact via a monopole-monopole force where, in the quasi-spin formulation, the Hamiltonian is given by
\begin{equation}\label{Eq:LMG}
	\hat{H} = \epsilon \hat{J}_z -\frac{V}{2} \left( \hat{J}_+^2 + \hat{J}_{-}^2\right) -\frac{W}{2} \left( \hat{J}_+\hat{J}_{-} + \hat{J}_{-}\hat{J}_+\right).
\end{equation}
The operators $\hat{J}_z$ and $\hat{J}_{\pm}$ are related to the fermionic field operators via
\be
{\hat J}_z = \frac{1}{2}\sum\limits_{m=1}^N\sum\limits_{\sigma=\pm } \sigma{\hat\psi}^{\dagger}_{\sigma m}{\hat\psi}_{\sigma m}, \ \ \ \ \ \ \ 
{\hat J}_{\sigma} =\sum\limits_{m=1}^N {\hat\psi}^{\dagger}_{\sigma m}{\hat\psi}_{-\sigma m}
\ee
and satisfy the angular momentum commutation algebra, while the index $\sigma=\pm $ differentiates the upper and lower levels. The interaction term associated with $V$ scatters two particles from the same energy level up or down, and $W$ scatters one particle up and another down or vice versa from different energy levels. The wavefunctions of the system can be expressed via the eigenstates $\ket{J, M}$ of the operators $\hat{J}_z$ and $\hat{J}^2 = \frac{1}{2} \lbrace \hat{J}_{+}, \hat{J}_{-} \rbrace + \hat{J}_z^2$, and  $\ket{J, M}$ serves as a convenient basis. These eigenstates are labeled by the quantum numbers ${\bf J} = {\bf j}_1 + {\bf j}_2 + \ldots + {\bf j}_N$, which is the total spin, and its projection $M$ in the $z$-direction. Symmetries of this model can be exploited to significantly reduce the size of the relevant Hilbert space. The first symmetry arises from the invariance of the Hamiltonian under the exchange of particles within the set of two levels. Additionally, in the case of $W=0$ in Eq. (\ref{Eq:LMG}), the interaction term only couples states that differ by spin $M\pm 2$; hence we can block-diagonalize the Hamiltonian. This leads to the maximally efficient encoding scheme (\textbf{J}-scheme) introduced in Ref. \cite{Hlatshwayo2022}. In this encoding scheme, the problem of finding eigenvalues of Eq. (\ref{Eq:LMG}) reduces to the diagonalization of smaller matrices of dimensions $J$ and $(J+1)$, 
where $J=\frac{1}{2}N$. Therefore, with the condition $W=0$, the Lipkin model has an $\mathcal{O}(N)$ complexity for diagonalizing its Hamiltonian, which enables quite efficient and noise-resilient quantum computation. 

The exact analytical solution of the LMG model with $W=0$ for systems with a few particles is given in Refs.~\cite{lipkin1965validity, co2015hartree}. Some extensions of the Lipkin model have been proposed, such as the Agassi model~\cite{agassi1968validity, perez2022digital}, the three-level Lipkin model \cite{xu1995development}, and the generalized Lipkin model~\cite{carrasco2016generalized}, all of which could be interesting test-beds for quantum algorithms.

\subsection{\label{sec:encode} Efficient Encoding}
\begin{figure}[h!]
\centering
    \begin{quantikz}[thin lines]
    \lstick{$\ket{0}$} & \gate{R_y(\phi_{0})}  & \qw &  \qw  & \qw \\
    \lstick{$\ket{0}$} & \gate{R_y(\phi_{1})} & \ctrl{-1}  & \gate{R_y(\phi_{2})} & \qw 
    \end{quantikz}
\caption{Ansatz circuit for block B of the Lipkin system with N=8 particles.}
\label{fig:N8_ansatz_B}
\end{figure}
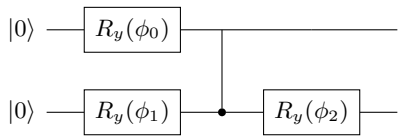

In Ref. \cite{Hlatshwayo2022}, we introduced the most efficient encoding scheme of the LMG Hamiltonian (\ref{Eq:LMG}) with $W=0$ on a quantum computer. This section briefly describes this encoding scheme dubbed as the \textbf{J}-scheme. As mentioned above, the multiplet representation $\ket{J, M}$ of the basis states helps reveal and exploit symmetries of the Hamiltonian. Each block contains a ladder of states with spin projections that differ by an even number of units. The first block can be mapped to qubits as follows
\begin{equation}\label{eq_SB_map}
\begin{aligned}
|J,-J\rangle & \equiv|0\rangle \rightarrow \ket{\text{bin}(0)}, \\
|J,-J+2\rangle & \equiv|1\rangle \rightarrow \ket{\text{bin}(1)},\\
\cdots & \\
|J,J-2\rangle & \equiv\left|d-2\right\rangle \rightarrow \ket{\text{bin}(d-2)}, \\
|J,J\rangle & \equiv\left|d-1\right\rangle \rightarrow \ket{\text{bin}(d-1)},
\end{aligned}
\end{equation}
where the Gray code (GC) is employed for $\ket{\text{bin}(k)}$, and the mapping of the other block is done similarly. The GC orders the binary values such that any two adjacent entries differ by only a single bit \cite{gray1953pulse}. The GC has been shown to be more efficient than standard binary coding for Hamiltonian simulations on a quantum computer ~\cite{sawaya2020resource,di2021improving}. Thus, the dimensionless Hamiltonian $\Bar{H}=\Hat{H}/\epsilon$ is recast as 
\begin{equation}\label{Eq:LMG_J}
    \Bar{H} = \sum_{k=0}^{d-1} a_k \ket{k}\bra{k} +  \sum_{k=0}^{d-2} b_k \biggl( \ket{k}\bra{k+1} + \ket{k+1}\bra{k} \biggr), 
\end{equation}
where the states $\ket{k}$ are mapped to $\ket{\text{bin}(k)}$ by the GC and the coefficients are given by
\begin{equation}\label{eq_ak}
    a_k =  M,
\end{equation}
\begin{equation}\label{eq_bk}
\begin{split}
   b_k &= -\frac{v}{2} \times \lbrace  \left[J(J+1) -M(M + 1)\right] \\
		&\times \left[J(J+1) - (M + 1)(M + 2)\right] \rbrace^{\frac{1}{2}}. 
\end{split}
\end{equation}
where $M=2k - J$ and $v=V/\epsilon$. In this work, we consider a system of $N=8$ particles, thus $J=4$, which corresponds to a total multiplet of 9 states and decomposes into two disconnected sub-blocks of even and odd values of $M$ denoted by A and B:  
\begin{equation}\label{Eq:GC_map_N8}
\begin{aligned}
\ket{4, -4} & \equiv \ket{0}_A \rightarrow \ket{000} \qquad & \ket{4, -3}  & \equiv \ket{0}_B \rightarrow \ket{00} \\
\ket{4, -2} & \equiv \ket{1}_A \rightarrow \ket{00{\bf 1}} \qquad & \ket{4, -1}  & \equiv \ket{1}_B  \rightarrow \ket{0{\bf 1}} \\
\ket{4, 0} & \equiv \ket{2}_A \rightarrow \ket{0{\bf 1}1} \qquad & \ket{4, +1}  & \equiv \ket{2}_B  \rightarrow \ket{{\bf 1}1}  \\
\ket{4, +2} & \equiv \ket{3}_A \rightarrow \ket{01{\bf 0}} \qquad & \ket{4, +3}  & \equiv \ket{3}_B \rightarrow \ket{1{\bf 0}}  \\
\ket{4, +4} & \equiv \ket{4}_A \rightarrow \ket{{\bf 1}10} \qquad &   
\end{aligned}  
\end{equation}
where the GC single bit that changes in subsequent states is shown in bold. The Hamiltonian for block B is given by the ansatz
\begin{equation}\label{Eq:ham_blockB}
\begin{split}
\Bar{H}_{B} &= a_0 \ket{00}\bra{00} + a_1 \ket{01}\bra{01} + a_2  \ket{11}\bra{11} + a_3 \ket{10}\bra{10} \\
&+ b_0 \biggl( \ket{00}\bra{01} + \ket{01}\bra{00} \biggr) + b_1 \biggl( \ket{01}\bra{11} + \ket{11}\bra{01} \biggr)\\
&+ b_2 \biggl( \ket{11}\bra{10} + \ket{10}\bra{11} \biggr),
\end{split}
\end{equation}
with the following coefficients:
\begin{equation}\label{Eq:hamA_coeff}
\begin{aligned}
  & a_0 = -3,  \quad a_1 = -1, \quad a_2 = 1, \quad a_3 = 3, \\
  & b_0 = - 3\sqrt{7}v, \quad b_1 = -10v, \quad b_2 = - 3\sqrt{7}v. 
\end{aligned}
\end{equation}
The Hamiltonian of Eq. (\ref{Eq:ham_blockB}) is then rewritten in terms of Pauli matrices by noting that the operators associated with $a_k$ are given by
\begin{equation}\label{Eq:ak_ops}
\begin{aligned}
\ket{00}\bra{00} &= P^{(0)}_1 P^{(0)}_0 = \frac{1}{4}\left( \mathbb{I} + Z_0 + Z_1 + Z_1 Z_0 \right) \, ,  \\
\ket{01}\bra{01} &= P^{(0)}_1 P^{(1)}_0 = \frac{1}{4}\left( \mathbb{I} - Z_0 + Z_1 - Z_1 Z_0 \right) \, ,  \\
\ket{11}\bra{11} &= P^{(1)}_1 P^{(1)}_0 = \frac{1}{4}\left( \mathbb{I} - Z_0 - Z_1 + Z_1 Z_0 \right) \, , \\ 
\ket{10}\bra{10} &= P^{(1)}_1 P^{(0)}_0 = \frac{1}{4}\left( \mathbb{I} + Z_0 - Z_1 - Z_1 Z_0 \right) \, ,  \\
\end{aligned} 
\end{equation}
where $P^{(0)}_i=\frac{1}{2}\left( \mathbb{I}_i + Z_i \right)$ and $P^{(1)}_i=\frac{1}{2}\left( \mathbb{I}_i - Z_i \right)$ are the projection operators acting on the $i^{\text{th}}$ qubit. Similarly, the operators associated with $b_k$ can be expressed as 
\begin{equation}\label{Eq:bk_ops}
\begin{aligned}
\ket{00}\bra{01}+ \ket{01}\bra{00}  &= P^{(0)}_1 X_0 = \frac{1}{2}\left( X_0 + Z_1 X_0 \right) \, , \\
\ket{01}\bra{11}+ \ket{11}\bra{01}  &= X_1 P^{(1)}_0 = \frac{1}{2}\left( X_1 - X_1 Z_0 \right) \, , \\
\ket{11}\bra{10} + \ket{10}\bra{11}  &=  P^{(1)}_1 X_0 = \frac{1}{2}\left( X_0 - Z_1 X_0 \right) \, .\\
\end{aligned}
\end{equation}
Note that the order of operations is important. For instance, the gate $Z_0$ should be interpreted as $\mathbb{I}_1 Z_0$ while $Z_1$ is $Z_1 \mathbb{I}_0$, so that the tensor product generates the proper matrix form. 
After substituting  Eqs. (\ref{Eq:hamA_coeff} - \ref{Eq:bk_ops}) into Eq. (\ref{Eq:ham_blockB}), the Hamiltonian of block B reads
\begin{equation}\label{Eq:ham_blockB_simplified}
\Bar{H}_{B} = -2Z_1 - Z_1Z_0 - 3\sqrt{7}v X_0 -5v X_1 + 5v X_1Z_0. 
\end{equation}
The associated GC wave function for block B is given by 
\begin{equation}
\begin{split}
\ket{\psi_B} &=\cos{\phi_0}\ket{00} + \sin{\phi_0}\cos{\phi_1}\ket{01} \\
&+ \sin{\phi_0}\sin{\phi_1}\cos{\phi_2}\ket{11} \\
&+\sin{\phi_0}\sin{\phi_1}\sin{\phi_2}\ket{10},
\end{split}
\end{equation}
which can be represented by the ansatz circuit shown in Fig. \ref{fig:N8_ansatz_B}.

\begin{figure}[h!]
\centering
    \begin{quantikz}[thin lines]
    \lstick{$\ket{0}$} & \gate{R_y(\phi_{0})}  & \qw &  \qw  & \qw & \qw \\
    \lstick{$\ket{0}$} & \gate{R_y(\phi_{1})} & \ctrl{-1}  & \qw & \qw  & \qw \\
    \lstick{$\ket{0}$} & \gate{R_y(\phi_{2})} & \qw  & \ctrl{-1} & \gate{R_y(\phi_{3})} & \qw 
    \end{quantikz}
\caption{Ansatz circuit for block A of the Lipkin system with N=8 particles.}
\label{fig:N8_ansatz_A}
\end{figure}
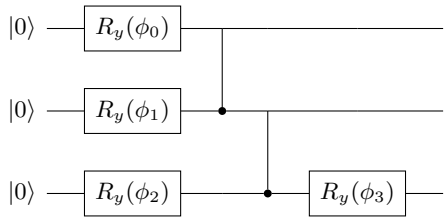

One can follow the same procedure to obtain the block-A Hamiltonian:
\begin{equation}
	\begin{split}
		\Bar{H}_A &=  -\left( IIZ + ZZI\right)  + \frac{1}{2} \left( ZII - 3IZI - IZZ - ZIZ \right)  \\
		&+ \frac{3\sqrt{10}}{4} v \left( ZXZ + ZXI -IXZ - IXI \right) \\
		&+ \frac{\sqrt{7}}{2} v \left( ZZX + IZX - ZIX - IIX \right) \\
		&-  \frac{v}{4} \left( 3\sqrt{10} + 2\sqrt{7} \right) \left(XII + XIZ \right) \\
		&+ \frac{v}{4} \left( 3\sqrt{10} - 2\sqrt{7} \right) \left( XZI + XZZ \right),
	\end{split}
\end{equation}
where the subscripts marking the qubits are dropped for readability. The associated wavefunction for block A can be represented by the ansatz circuit shown in Fig. \ref{fig:N8_ansatz_A}.

In general, for an arbitrary $N$, the Hamiltonian is split into block form as 
\begin{equation}\label{eq_gen_H}
	\Bar{H}^{(N)}_J  = \begin{pmatrix}
	\Bar{H}_A & 0 \\
	0 & \Bar{H}_B
	\end{pmatrix},
\end{equation}
where the dimension of block $A$ and $B$ is: $d_A = J+1$ and $d_B = J$ for the even values of $N$, and $d_A=d_B = \frac{1}{2}(N+1)$ for the odd values of $N$. In Ref. \cite{Hlatshwayo2022}, we showed that this encoding leads to a logarithmic scaling of the number of qubits $n_q$ with respect to the number of particles $N$ given by
\begin{equation}
    n_q = \lfloor \log{_2\left(\frac{N}{2}+1\right)} \rfloor.  
\end{equation}
This $\mathcal{O}(\log(N))$ scaling of the number of qubits is the critical factor that enables the realization of the quantum benefit. 

\subsection{Scaling of qEOM}
VQE+qEOM algorithm consists of four steps characterized by scaling with the number of particles $N$ in the system, namely:
\begin{enumerate}
	\item The number of qubits required to accommodate all the system states. 
	\item The number of independent parameters (angles) that express the many-body wavefunction of the system.
	\item The number of measurements of the basis Pauli-gates on a quantum computer required to evaluate the qEOM matrix elements.
	\item The dimension of the GEE solved on a classical computer.
\end{enumerate}
In this section, we determine the scaling of each algorithmic step for the Lipkin model. As mentioned in Sec {\ref{sec:encode}}, the number of qubits scales logarithmically with $N$. The number of angles $n_a$ is defined by
\begin{equation}
\begin{split}
n_a = 2^{n_q} - 1 = 2^{(\lfloor \log{_2\left(\frac{N}{2}+1\right)} \rfloor)} -1 \approx \frac{N}{2}.
\end{split}
\end{equation} 
Hence, the number of angles to be optimized for the wavefunction ansatz scales as $\mathcal{O}\left(N\right)$, i.e., linearly with $N$.

The total number of measurements $M$ needed to construct the qEOM matrix elements is determined by counting the combinations $\mathcal{P}_k$ of choosing $k$ Pauli strings from the set of four Pauli strings $\{ I, X, Y, Z \}$ including repetitions: 
\begin{equation}\label{Eq:permutations}
	M = 4^k - 1 = (2^{k})^2 - 1 \approx N^2,
\end{equation}
where $k = n_q$ is the number of qubits in the system. Therefore, since $k$ scales logarithmically with $N$, the number of measurements $M$ scales as $ \mathcal{O}(N^2)$, i.e., quadratically with $N$. In practice, not all the Pauli-strings in $\mathcal{P}_k$ enter the matrix elements; thus, the total measurements are much less than $N^2$, which represents the worst-case scenario. 
For the Lipkin model with $W=0$ the quadratic scaling makes the quantum benefit accessible by high-complexity approximations for strongly coupled systems. 

Finally, the scaling of the GEE matrix dimensions is found by computing the total number of unique excitations on an $n_q$-qubit quantum computer. According to the set theory, the total number of distinct proper subsets of the set $\lbrace 0, 1, \ldots q-1\rbrace$ is given by  $s=2^{n_q} -1$. We note that a proper subset excludes the original set but includes the null set $\lbrace \emptyset \rbrace$, whereas, in the case of unique excitations, we include the original set and exclude the null set; hence, the count is the same. The dimension of the GEE matrices is given by
\begin{equation}
    2s = 2\times (2^{n_q} -1) \approx N.
\end{equation}
Thus the GEE matrix dimension scales linearly as $\mathcal{O}(N)$ for the Lipkin model with the efficient \textbf{J}-scheme.

In general, assuming no symmetries can be exploited in the model Hamiltonian, the number of Pauli-gate measurements on a quantum computer and the dimension of the GEE solved by the classical computer grows exponentially with the number of particles in the system. 
However, in practical implementations of the EOM, and in nuclear structure in particular, the particle-hole pairs in the operators $\hat F$ and ${\hat Q}^{\dagger}$ are coupled either to a good angular momentum (in the spherical symmetry) or to a total angular momentum projection (in the axial symmetry) \cite{suhonen2007}. Accordingly, separate EOMs are solved for each set of quantum numbers, such as spin, isospin, and parity, which define an eigenmode or are transferred by an external field. This is an essential factor that moderates the scaling considerably.

\section{\label{sec:results} Quantum simulation results}

\begin{figure}
	\includegraphics[width=\linewidth]{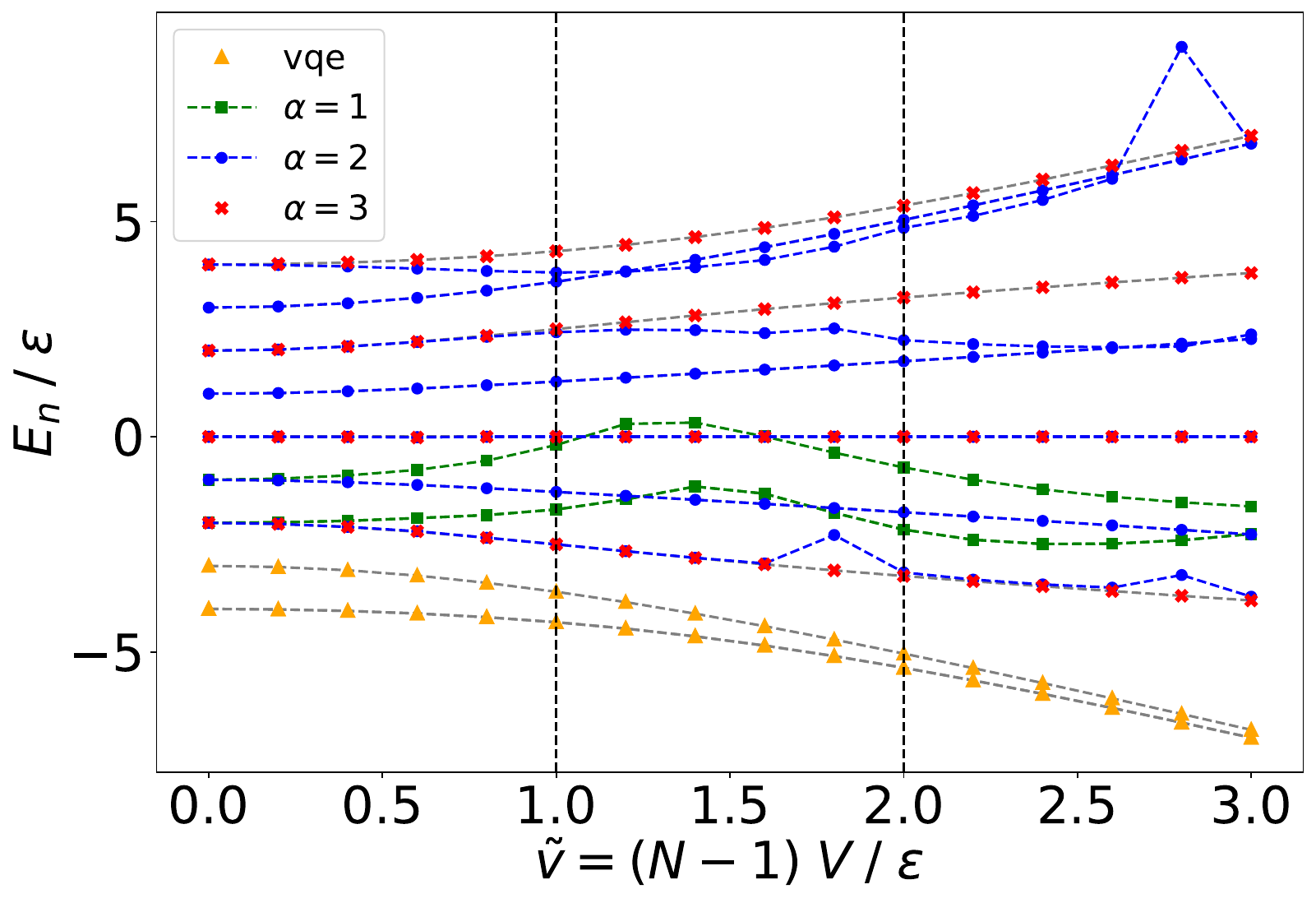}
 \caption{Simulator results for the energy spectrum of the Lipkin model as a function of the effective interaction strength $\tilde{v}$ for a system of $N=8$ particles. 
 The square, circle, and cross symbols represent calculations with configuration complexity $\alpha = 1, 2$ and $3$. The dashed lines are the exact solutions for the energy levels.}
\label{fig:En_spectrum_sim_dense}
\end{figure}

The energy spectrum of the Lipkin Hamiltonian (\ref{Eq:LMG}) with $W = 0$ and $N = 8$
is displayed in Fig. \ref{fig:En_spectrum_sim_dense}. The spectrum was obtained using a classical simulation of an ideal quantum computer with no noise errors, often called a simulator, with the running effective interaction strength $\tilde{v}=(N-1)v$. The results obtained on the simulator served as
a benchmark for the real noisy quantum device calculations. 
The ground $E_0$ and first excited $E_1$ states were found via the VQE minimization procedure for blocks A and B, respectively. The higher-energy states $\{E_2, E_4, E_6, E_8\}$ and $\{E_3, E_5, E_7\}$ are obtained by accordingly applying the excitation operator (\ref{Eqn:excitation_op}) on the two lowest-energy states.
The excited state energies 
of block A are found using the qEOM algorithm with $\alpha = 1$, $2$ and $3$ configuration complexities of the excitation operator, while the energies 
of block B were obtained with $\alpha = 1$, $2$.  
In the efficient J-scheme encoding of the Lipkin model, the maximal configuration complexity is $\alpha_m = n_q$, where $n_q$ is the number of qubits in the ansatz. As $n_q = 3$ for block A and $n_q =2$ for block B, all the possibilities were realized in the calculations. 
 The approximations to the excitation operator with the maximal configuration complexities $\alpha_m = 1$, 2, and 3 are dubbed as $\alpha_1$, $\alpha_2$ and $\alpha_3$, respectively. As the ground state is obtained by minimization and accounts for many-body correlations nearly exactly, these approximations differ from the regular RPA, second RPA, and third RPA based on the quasiboson approximation \cite{RingSchuck1980}. Therefore, the maximal configuration complexity index $\alpha_m$ is used for identification of the approach to the excitation operator.

\begin{figure}
	\includegraphics[width=\linewidth]{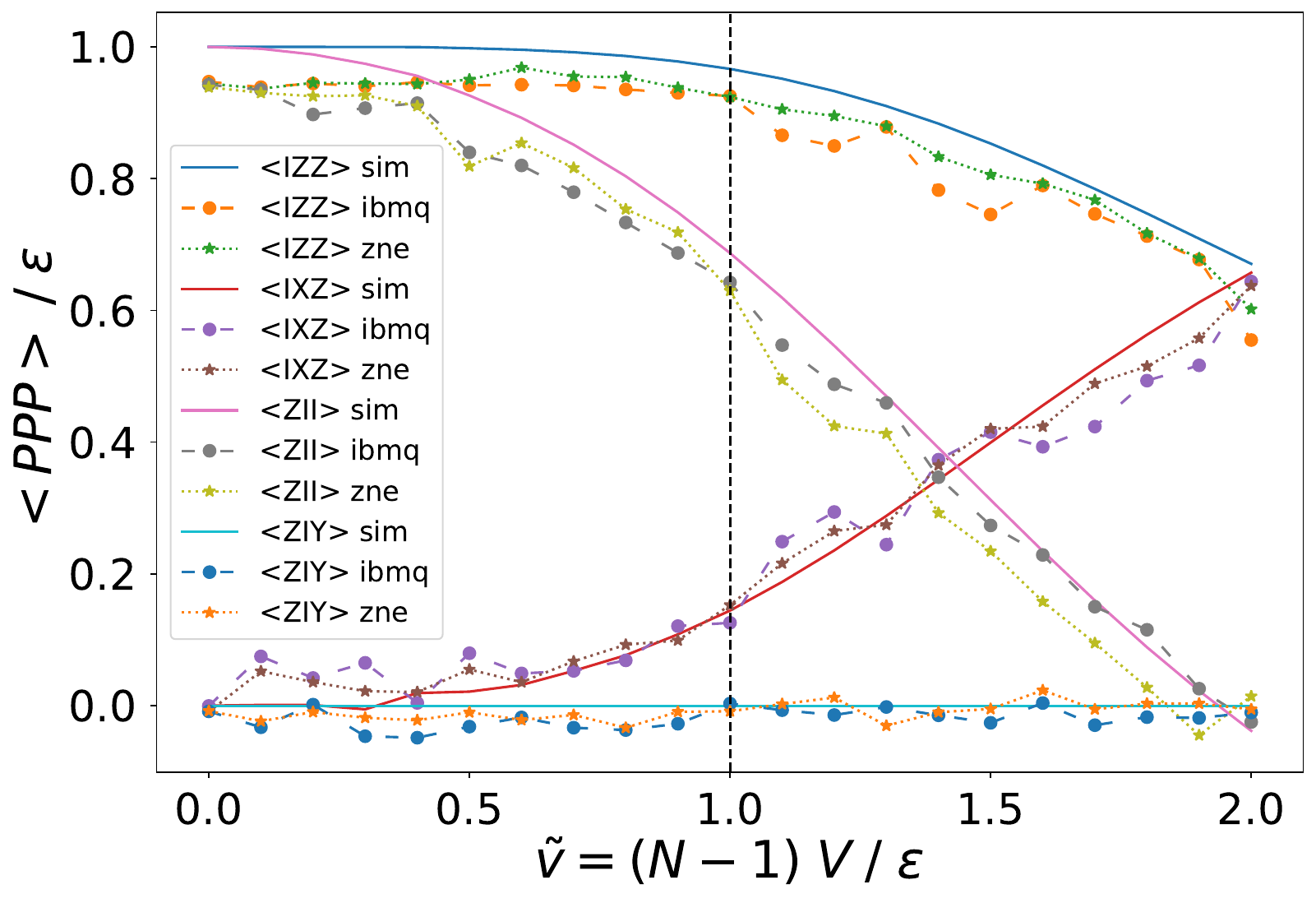}
 \caption{Basis expectation values as a function of the effective interaction strength for a three-qubits ansatz of the Lipkin model. The graph labels "ibmq" and "zne" represent computation performed on IBM quantum computers without error mitigation and with error mitigation employing the zero-noise extrapolation (ZNE) method, respectively. The solid lines labeled "sim" represent the values computed by the statevector simulator. }
\label{fig:Expectation_vals}
\end{figure}

\begin{figure}
	\includegraphics[width=\linewidth]{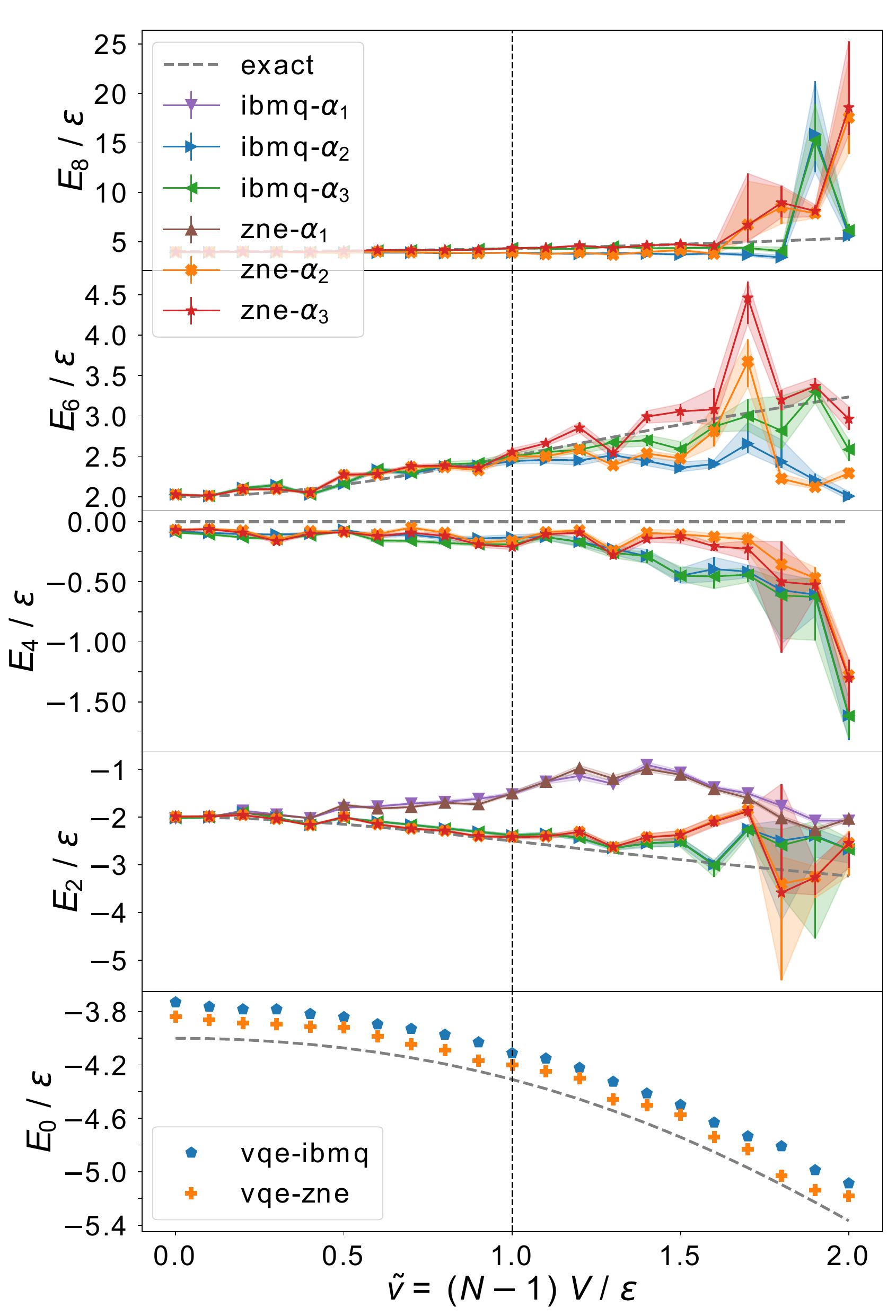}
	\caption{Energy spectrum of the Lipkin model as a function of the effective interaction strength $\tilde{v}$ for a system of $N=8$ particles. Energy levels of the Hamiltonian matrix block A are shown. The graph labels "ibmq" and "zne" represent computation performed on IBM quantum computers without error mitigation and with error mitigation employing the zero-noise extrapolation (ZNE) method, respectively.
 }
	\label{fig:En_spectrum_N8_A}
\end{figure}

We are particularly interested in how the increase in configuration complexity influences the results across coupling regimes. By construction of the excitation operator, it is expected that (i) larger $\alpha_m$ should lead to a more accurate solution and (ii) terms with larger $\alpha$ should become increasingly important with the growing interaction strength.
Indeed, one can see in Fig. \ref{fig:En_spectrum_sim_dense} that both trends manifest in both blocks A and B of the model Hamiltonian.

Because of the presence of the ground state correlations, the $\alpha_1$ solutions avoid the anomalous behavior of RPA at $\tilde v$ = 1 and rather resemble the pattern of the self-consistent RPA \cite{SchuckTohyama2016a,Schuck2021}. With the $\tilde v$ increase, the $\alpha_1$ solutions exhibit large deviations from the exact ones.
Introducing the $\alpha$ = 2 configurations into the excitation operator enables considerably more accurate solutions for all the energy levels. The $\alpha_m$ = 2 approximation demonstrates a very good performance for the $E_2 - E_5$ and $E_7$ excited states for almost all interaction strength values except for a few deviations in $E_2$ from the exact energies at strong coupling. The higher-energy $\alpha_m$ = 2 block-A solutions $E_6$ and $E_8$ start to deteriorate already near $\tilde v = 1$. Finally, with the maximal possible configuration complexity
$\alpha$ = 3 for block A, one can see a significant improvement of the results approaching the exact values. Thus, maximal possible configuration complexities generate nearly exact results with the VQE+qEOM algorithm on the simulator.

\begin{figure}
	\includegraphics[width=\linewidth]{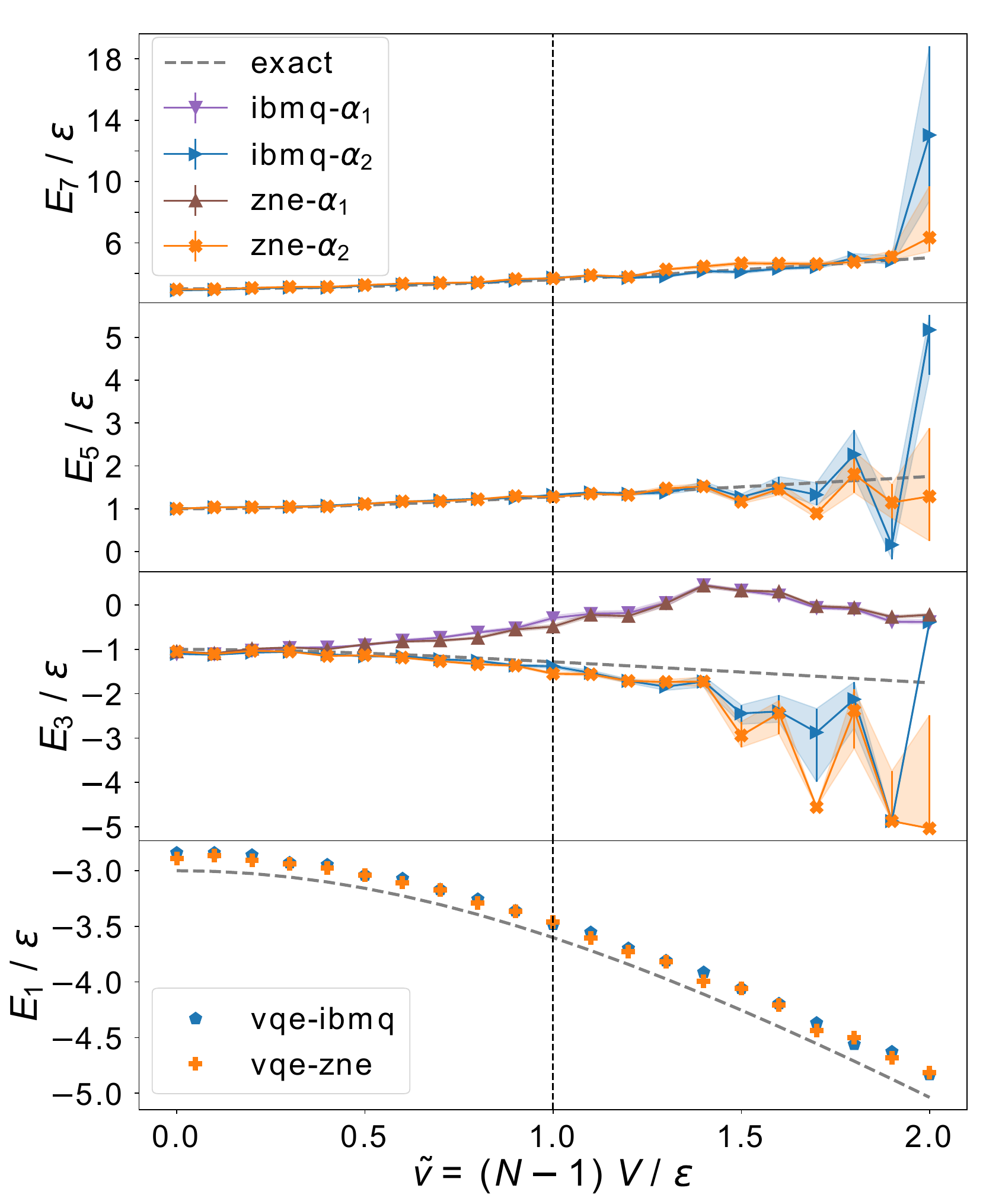}
 \caption{Same as in Fig. \ref{fig:En_spectrum_N8_A} but for block B of the Hamiltonian matrix.}
	\label{fig:En_spectrum_N8_B}
\end{figure}

Although the same trends manifest in the solutions on the quantum computer, the hardware results are affected by noise, which is especially pronounced at strong coupling. 
To reduce systematic errors from noise, zero-noise extrapolation (ZNE) error mitigation was adopted from Refs. \cite{ZNE2017,ZNE2017a} and applied to the final expectation value measurements.
ZNE consists of (i) amplifying the noise in the circuit by creating multiple circuit copies with varying multiplication factors of the $U^{\dagger}U$ type or local folding of single gates, (ii) making measurements for the multiplied circuits, thereby generating varying levels of noise, and (iii) using the results of the measurements to extrapolate down to the ideal setting with theoretically zero noise. 

The effect of the ZNE application on the single Pauli strings is illustrated in Fig. \ref{fig:Expectation_vals} for the selected elements $\langle IZZ \rangle$,  $\langle IXZ \rangle$, $\langle ZII \rangle$ and $\langle ZIY \rangle$ of the block-A computation. The hardware results before and after the error mitigation are displayed in comparison with the simulator results.  
The expectation values of the first two Pauli strings are systematically underestimated, while somewhat oscillating behavior is observed for the other two. Remarkably, the errors generally do not increase with the $\tilde v$ value, i.e., the strong coupling is not problematic at this step.  
The major trend in the ZNE results is a systematic reduction of hardware errors. One case where the ZNE does not improve the result is the strong coupling regime with $1.0 \leq \tilde v \leq 2.0$ for the $ZII$ string; however, the corresponding hardware values are quite close to the ones obtained on the simulator, i.e., not much affected by the noise. This may be a feature of this particular combination of the Pauli gates with two unit elements less affected by the noise.      

The IBM hardware results obtained by the same algorithm are demonstrated in  Figs. \ref{fig:En_spectrum_N8_A} and \ref{fig:En_spectrum_N8_B} in comparison with the exact solution. 
Fig. \ref{fig:En_spectrum_N8_A} shows the even energy levels of block A of the Lipkin Hamiltonian, and Fig. \ref{fig:En_spectrum_N8_B} displays the odd energy levels from its block B. 
 The results were collected before and after the  ZNE was applied to the basis Pauli-gates expectation value measurements and augmented with error bars determined as described in Section \ref{appendix:errorsadd}. 
The results before and after applying ZNE are marked by "ibmq" and "zne", respectively, and include computation with varying maximal configuration complexity $\alpha_m$ of the excited states. The ground state and first excited state energies are the outputs of the VQE for block A and block B, correspondingly. One can observe systematic inaccuracies in the VQE calculations, which return slightly larger energy values regardless of the interaction strength. While consistency and stability of VQE are discussed in Section \ref{appendix:errorsadd}, here we note that the application of ZNE definitely helps reduce the inaccuracies in the ground state energy in all the coupling regimes. However, even after the error mitigation, the ground state description is imperfect, so one can expect propagation of the remaining errors to the expectation values and, therefore, their non-negligible contribution to the errors in the excitation energies.

 The overall observation across the LMG spectrum is that, for the excited states, the hardware noise generates systematic errors following the same trends as the accuracy: the errors are amplified with the increase of $\alpha_m$ and growth of the interaction strength. As noted above, the hardware errors of single quantum measurements do not grow with the interaction strength; however, the role of larger $\alpha$ increases with stronger coupling, while the increase of $\alpha_m$ leads to more terms in the GEE matrix elements and eventually to post-processing higher-rank matrices.  These two factors together, therefore, lead to larger errors in the resulting energy spectrum at large $\tilde v$.  Overall, in the strong coupling regime, the reduction of theoretical errors by introducing higher complexity terms in the excitation operator comes at the price of amplified hardware errors. 

In most cases, the ZNE allows for a significant reduction in the errors originating from the hardware noise, according to the trend for the single Pauli strings. However, as follows from the sampling noise analysis discussed in Section \ref{appendix:errorsadd}, sampling errors in certain coupling regimes may dominate over the hardware errors.

\section{\label{appendix:errorsadd} Error analysis} 


Two sources of error were considered and investigated for this analysis: inaccuracies in the parameterization of the ground state wave function and noise from the quantum computer. The primary source of the former is inconsistent convergence of the VQE, while the latter arises from the probabilistic nature of the computation and noise in NISQ devices.
To get an approximate confidence interval for the expectation value measurements on a quantum computer, we employ a simple statistical model assuming that sampling variances are the dominant source of errors. 
\begin{figure}
	\includegraphics[width=\columnwidth]{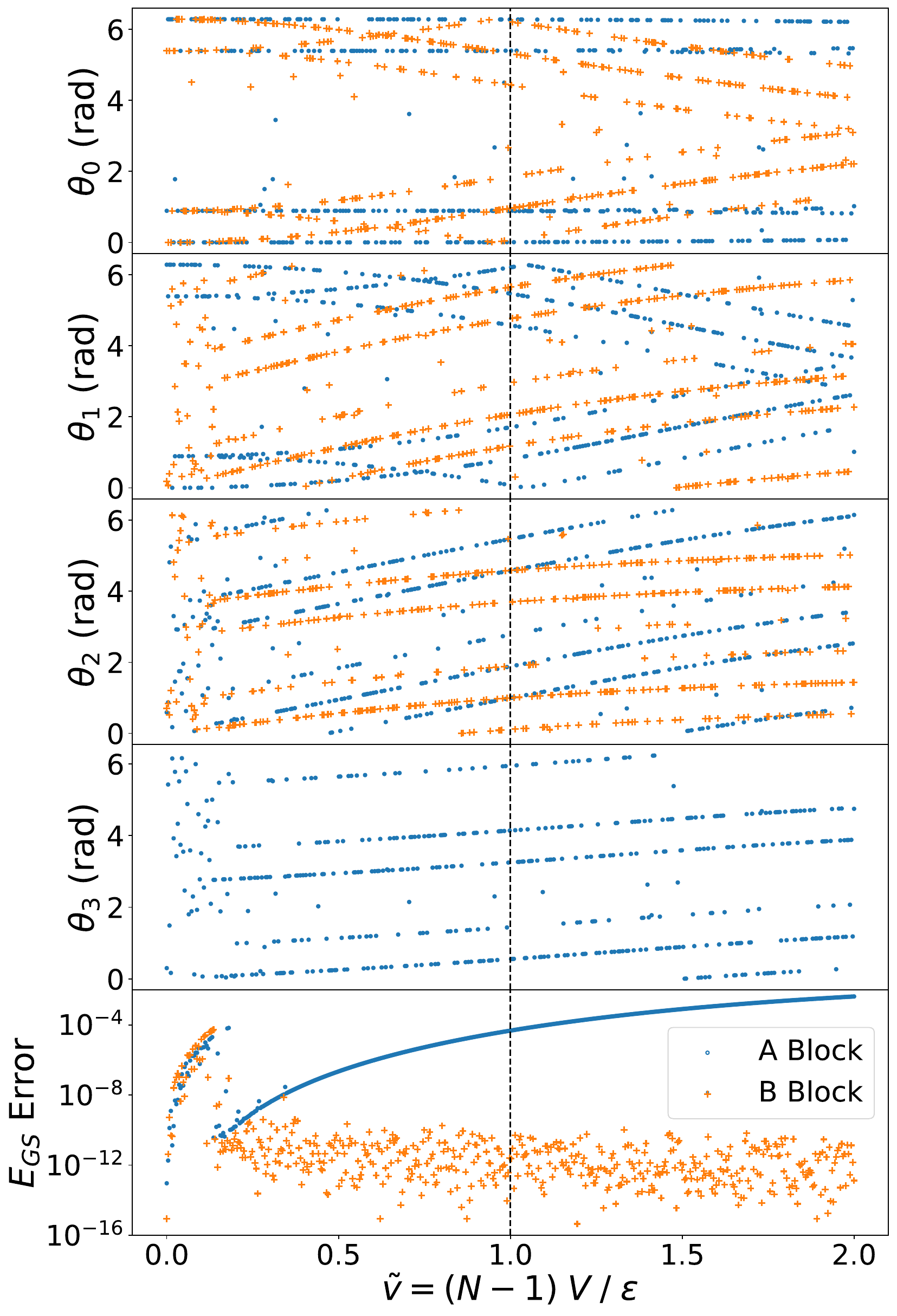}
	\cprotect\caption{Angles parameterizing the ground state wave functions. The top four panels display the value of the angles returned by VQE as a function of $\tilde{v}$ for blocks A and B. The bottom panel shows the error in the ground state energy ($E_0$) and the first excited state ($E_1$, the ground state of block B) calculated using the corresponding angles. See text for details.}
	\label{fig:VQE_angles}
\end{figure}

\subsection{\label{sec:VQE_stability} VQE Consistency and Stability}

VQE is a stochastic algorithm \cite{Tilly2022}, and there is neither a guarantee that it will converge to the global energy minima nor that repeated iterations will converge to the same local minima. In practice, it was observed that repeated iterations of VQE converged to a consistent set of solutions that varied smoothly with $\tilde v$. Fig. \ref{fig:VQE_angles} illustrates the "optimal" angles generated by VQE for 500 uniformly distributed values of $\tilde{v}$ from $0.0$ to $3.0$. As is shown in Fig. \ref{fig:N8_ansatz_A}, the ground state wave function of block A of the Lipkin model for $N=8$ is parameterized by 4 angles (labeled $\theta_i; i\in[0, 1, 2, 3]$), and block B by 3 angles (labeled $\theta_i; i\in[0, 1, 2]$, see Fig. \ref{fig:N8_ansatz_B}). Adjacent values of $\tilde{v}$ are not necessarily continuous, but a band structure is visible. In practice, it was found that these different sets of optimal angles generated effectively identical energy spectra 
and so are presumably equivalent. Ground state energies were calculated directly, on a simulator, from the angles in  Fig. \ref{fig:VQE_angles} and found to agree with exact solutions within $10^{-4}$. Likewise, a separate analysis was performed in simulation for 25 uniformly distributed values of effective interaction strength, where the energy spectra were calculated 50 times using a different set of "optimal" angles generated by VQE. The resulting energy spectra were consistent within numerical precision.

\begin{figure}
	\includegraphics[width=\columnwidth]{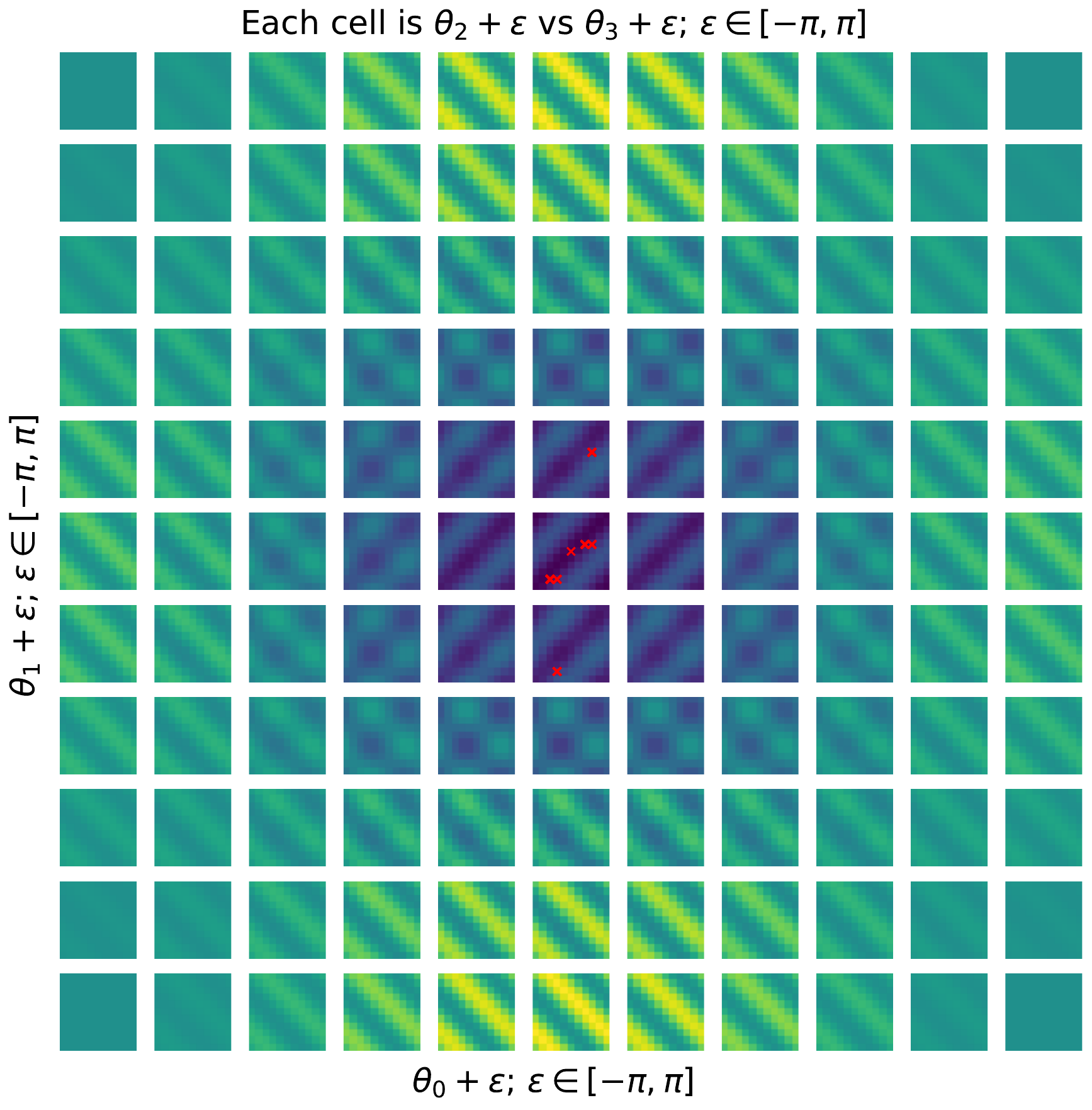}
	\cprotect\caption{Ground state energy of Lipkin model with $N=8$ as a function of block A parametrization computed on a simulator. Dark blue corresponds to lower energy, and bright yellow to higher energy. Each row of the grid is a single value of $\theta_1$, and the value is incremented continuously for each row. Each column of the gird is a single value of $\theta_0$, and the value is incremented for each column. Each cell of the grid is a heat map with $\theta_2$ on the x-axis and $\theta_3$ on the y-axis. The angles have all been shifted so that the center of the figure corresponds to an arbitrarily chosen set of optimal angles from VQE. The red X markers indicate specific sets of angles found by VQE, rounded to the nearest values used for the sampling. Note that $\varepsilon$ represents the angle variation and is not to be confused with the energy scale of the Lipkin model.
 }
	\label{fig:VQE_GS_GOH}
\end{figure}

The presence of multiple sets of optimal angles and the consistency of their quality suggests that there exist multiple minima in the VQE parameter space that are effectively equivalent. To study this, the ground state energy was calculated across the entire range of the angles parameterizing block A with $\tilde{v}=0$. A discrete uniform sampling of 10 steps was taken for each of the four angles from $-\pi$ to $\pi$, and the ground state was calculated using a simulator. Fig. \ref{fig:VQE_GS_GOH} illustrates this analysis, as well as sets of angles returned by VQE. The same analysis was performed for $\tilde{v} > 0$, and the results were materially similar but with the locations of the minima shifted corresponding to the shift in the band structure seen in Fig. \ref{fig:VQE_angles}. The results reported in Fig. \ref{fig:VQE_GS_GOH} provide an important insight into the stability of VQE. The dark blue spots in the center of the plot represent a region of lower energy close to or equal to the ground state energy $E_0$. 
We observe that the radius of this region is large; that is, $\theta_0$ and $\theta_1$ can be changed by up to $\pi/2$ from the "optimal" values, and the energy will still be relatively close to $E_0$. This implies the VQE minimization, at least in this case, is robust against small changes in the optimal parameters. In other words, if the dark blue region were small such that small changes of $\varepsilon$ on the optimal angles resulted in energy solutions far from $E_0$ (in the light blue region), then VQE would be unstable and susceptible to the barren plateau problem.

\begin{figure}
	\includegraphics[width=\columnwidth]{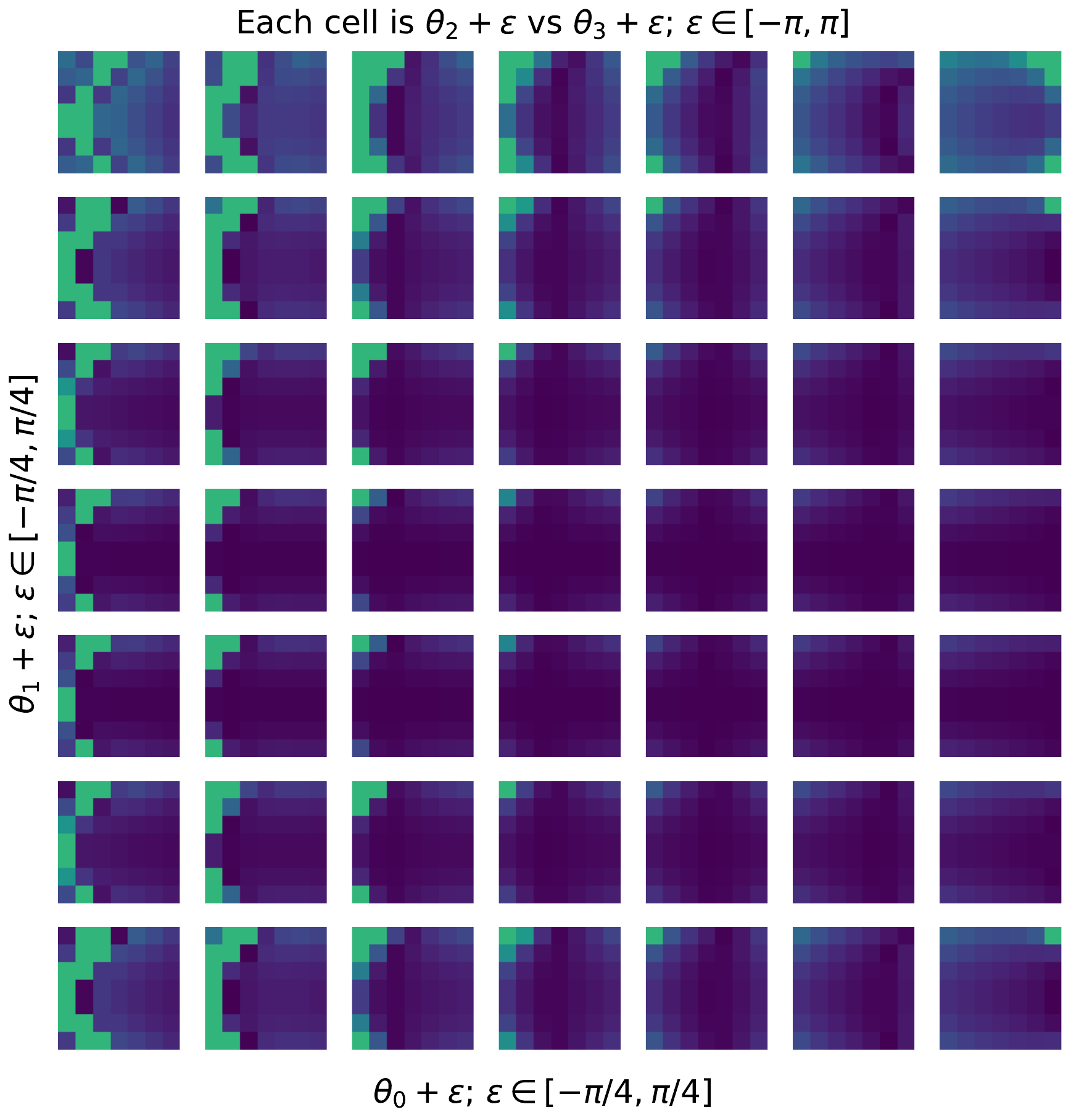}
	\cprotect\caption{Deviation of $E_7/\varepsilon$ from the exact solution for the Lipkin model with $N=8$, configuration complexity $\alpha=2$, and $\tilde{v}=0$. Each row of the grid is a single value of $\theta_1$, and the value is incremented continuously for each row. Each column of the gird is a single value of $\theta_0$, and the value is incremented for each column. Each cell of the grid is a heat map with $\theta_2$ on the x-axis and $\theta_3$ on the y-axis. The color scale is logarithmic with dark blue corresponding to a smaller deviation and bright yellow to a larger deviation. The minimum and maximum values of the color scale are the same as in Fig. \ref{fig:GOH_a3_E6_v1.7}. The angles have all been shifted so that the center of the figure corresponds to an arbitrarily chosen set of optimal angles from VQE. This figure corresponds to the second panel from the top in Fig. \ref{fig:spectra_w_err} at $\tilde{v}=0$.}
	\label{fig:GOH_a3_E6_v0.0}
\end{figure}

\begin{figure}
	\includegraphics[width=\columnwidth]{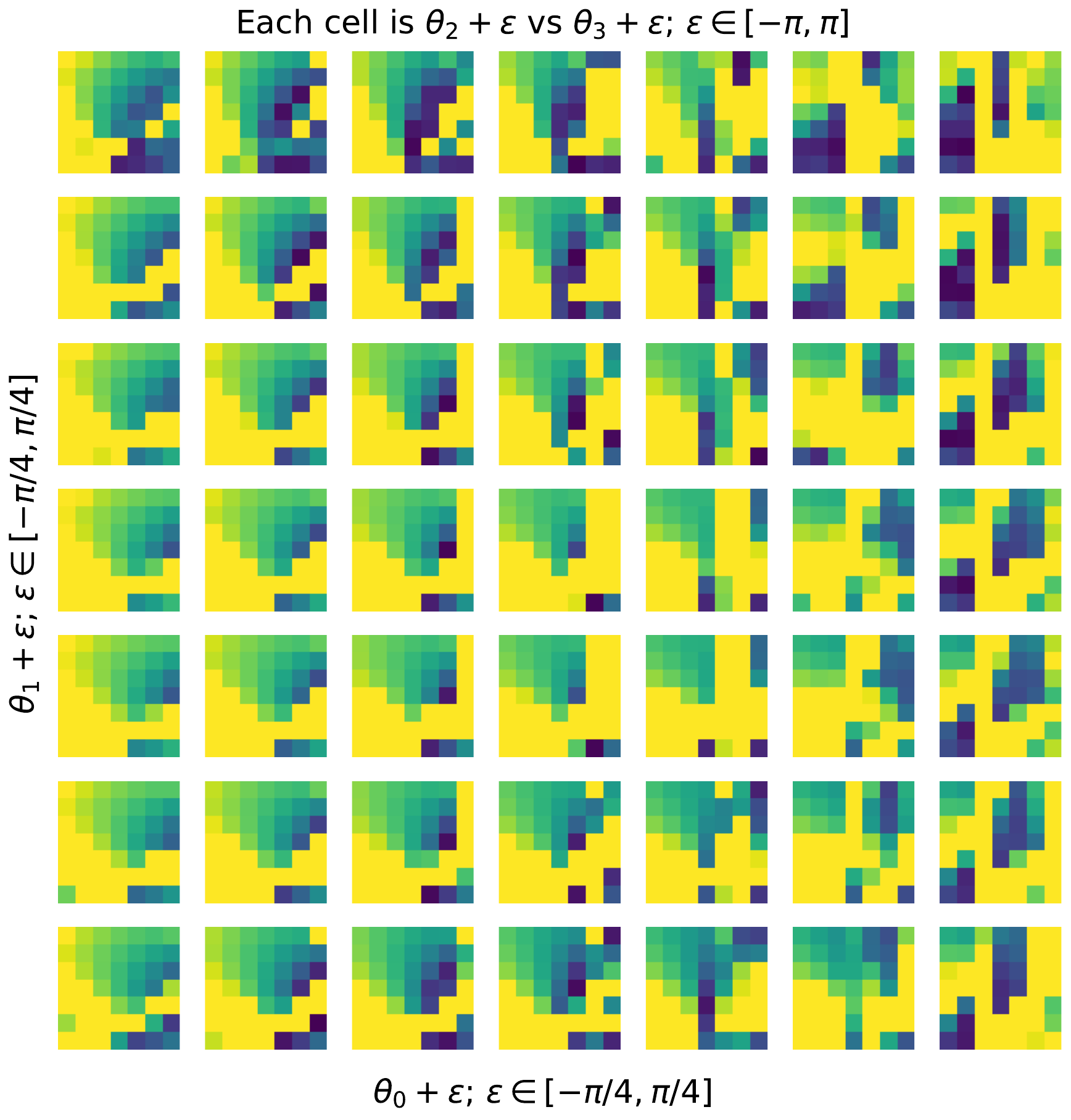}
	\cprotect\caption{Same as Fig. \ref{fig:GOH_a3_E6_v0.0} but with $\tilde{v}=1.7$. The minimum and maximum values of the color scale are the same as in Fig. \ref{fig:GOH_a3_E6_v0.0}. This figure corresponds to the second panel from the top in Fig. \ref{fig:spectra_w_err} where the error band gets large at $\tilde{v}=1.7$. }
	\label{fig:GOH_a3_E6_v1.7}
\end{figure}

A similar analysis was performed on the effect of small changes in the ground state parameterization on the excited state energies. The expectation values of the Pauli operators were calculated for one set of optimal angles found by VQE and kept constant for all subsequent calculations. The analysis was done using a discrete uniform sampling of 7 steps for $\theta_i; i\in \{0, 1, 2, 3\}$ of the ground state wave function of block A from $-\pi/4$ to $\pi/4$. The results of this analysis for $E_7$, configuration complexity $\alpha = 2$ for $\tilde{v}=0.0$ and $1.7$ are displayed in Fig. \ref{fig:GOH_a3_E6_v0.0} and Fig. \ref{fig:GOH_a3_E6_v1.7}, respectively. In general, the effect of small changes to the parameterization is small, indicating robustness against small misparameterization. However, in some cases (such as shown in Fig. \ref{fig:GOH_a3_E6_v1.7}), the energy spectra are highly sensitive to the ground state parameterization. These cases are associated with regions of high sensitivity to sampling noise, as discussed in Subsection \ref{appendix:sampling_noise}.

\subsection{\label{appendix:sampling_noise} Analysing noise from quantum measurements}

The probabilistic nature of quantum computation and the presence of stochastic noise in NISQ devices necessitates running all computations multiple times (known as the number of "shots") and calculating mean values. Broadly speaking, two types of error can be considered: sampling noise (arising from a finite number of shots) and systematic errors inherent in the nature of the hardware. The error bars in this work represent the uncertainty from sampling noise, and no attempt has been made to calculate error bars for systematic errors. As discussed above, ZNE demonstrates the ability to mitigate the effect of noise-induced systematic errors.


In this analysis, we conjecture that the individual measurements (i.e., the individual shots) can be taken from a binomial distribution (appropriately scaled). In practice, when considering the individual measurements, if the hardware is stable (and ignoring the complexities involved in performing and post-processing the measurements), there should be no difference between one run of 50,000 shots and 50 runs of 1,000 shots, as they are all coming from the same binomial distribution with the same expectation value. Given this observation, one can note two things: firstly, the sampling noise is a function of the expectation value and the number of samples taken, and secondly, it is trivial to simulate this sampling noise without the need for the quantum computer. This observation was verified for a subset of the measurements taken on the quantum computer. 

A more rigorous approach for modeling the noise would also consider error sources from (i) hardware infidelities in the
form of depolarizing Pauli noise, (ii) state preparation
and measurement (SPAM) errors, and (iii) decoherence
in the form of thermal relaxation and dephasing, see Refs. \cite{Georgopoulos2021,Wallman2016}. But for the Lipkin model with a few particles, we found that the reported variances from "ibmq'' are consistent with the binomial model, and the variance of the measured expectation values is consistent with our simple statistical model of the noise. Hence we computed error bars of the measured expectation values from a quantum computer using our simple model. 


The relationship between the noise from the quantum computer and the energy spectra is non-trivial because of the multi-step nature of the qEOM algorithm. In practice, it depends on the number of particles considered, the circuit built to accommodate the Hamiltonian, the embedding, and the eigensolver used. We assessed the effect of sampling noise on the energy spectra by calculating the spectra multiple times using different values of each measured operator. For the calculation considered, the GEE was solved 100 times using for each operator a value sampled from a binomial distribution with an expectation value given by the quantum computer (or the simulator) and a sample size of $2^{13}=8192$. The resulting distributions for the energy spectra are highly variable, depending on energy level, $\tilde{v}$, and configuration complexity $\alpha_m$ and, in most cases, highly asymmetric so the error bars presented are in the range from first to third quartiles.

\begin{figure}
	\includegraphics[width=\columnwidth]{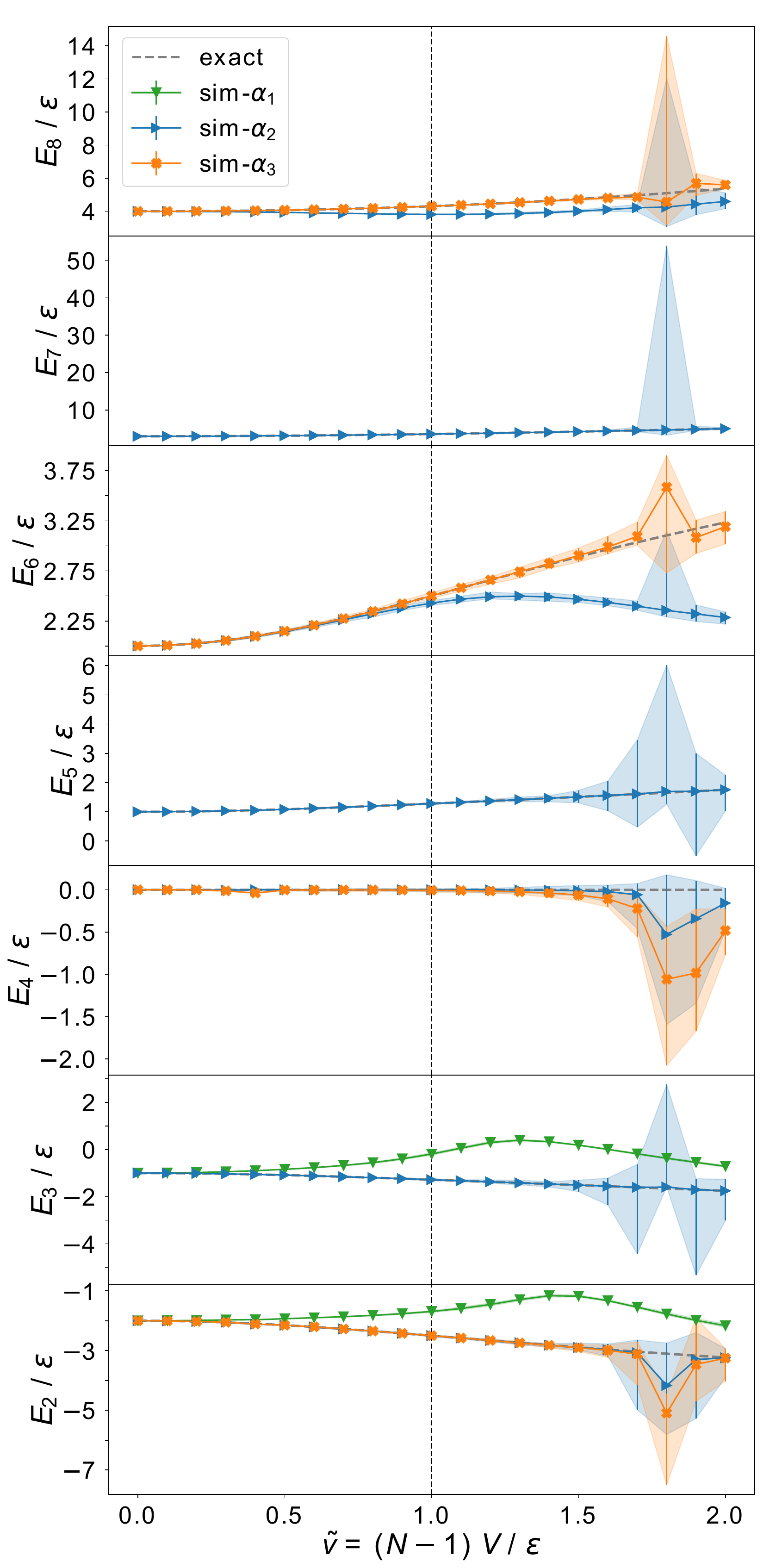}
	\cprotect\caption{Energy spectra of excited states for the Lipkin model with $N=8$ calculated using the emulator for configuration complexities $\alpha=1, 2$ and $3$. Error bands represent the effect of noise. See text for details.}
	\label{fig:spectra_w_err}
\end{figure}

To further verify the validity of the simple statistical noise model, we computed the excitation energies on a simulator with noise from our model, which, for clarity, we shall call an $\emph{emulator}$.   
Fig. \ref{fig:spectra_w_err} shows the emulator results for the excited states 
where the error bands represent the effect of noise, the marker corresponds to calculations without noise, and
%
the error bars represent the range from the first to third quartiles. 
Comparing the results of Fig. \ref{fig:spectra_w_err} with those shown in Figs. \ref{fig:En_spectrum_N8_A} and \ref{fig:En_spectrum_N8_B} suggests that the noise is strong in the regimes where large errors are observed, for instance, in the vicinity of $\tilde v = 1.7$. This is the same range of coupling strength where the VQE is highly sensitive to the sampling noise.

Furthermore, as the ground state $\ket{0}$ is computed by VQE, i.e., in a procedure separate from the qEOM for the excited states, the vacuum annihilation condition given by Eq. (\ref{vacuum}) is not necessarily satisfied in practice. The reason is that the ground state wave function and excitation operator are adopted with different correlation contents. This may potentially introduce theoretical errors; however, the self-consistency can be restored in a straightforward way. Some relevant methods are discussed in \cite{Datta1993,Szekeres2001}, 
and a unitary transformation of the excitation operator restoring the VAC is implemented in \cite{Asthana2023} in the framework of the quantum self-consistent EOM applied to molecular calculations. We leave an exploration of a self-consistent qEOM with strong coupling for future endeavors.

\section{\label{sec:summary} Summary and outlook }

In this work, we explored the potential of the VQE+qEOM method for quantum hardware computation of strongly coupled fermionic systems. Based on the maximally efficient encoding scheme \cite{Hlatshwayo2022}, the method was implemented for the Lipkin Hamiltonian across the regimes between weak and strong couplings and executed on NISQ devices. For the system of $N = 8$ fermions, the method allows generating excitations with up to $3p3h$ configuration complexity. While the quantum chemistry realm is confined by the weak coupling regime and thus qEOM provides an accurate description of electronic systems at the $2p2h$ level \cite{Ollitrault2020,Asthana2023}, in systems dominated by strong interactions higher configuration complexity is often needed for an adequate theoretical description of spectral phenomena. We show explicitly how higher-complexity configurations become increasingly important with the increase of the effective interaction strength, including the particle number scaling factor. This effect is related to the emergence of collective behavior of strongly coupled fermions.

The quantum benefit of the qEOM method was demonstrated. We found that increases in configuration complexity only increase the number of terms in the matrix elements of the generalized eigenvalue equation but do not affect the number of quantum measurements. The latter is fixed by the number of qubits employed in implementing the model Hamiltonian.
NISQ simulations on IBM quantum computers confirmed the robustness of the algorithm and demonstrated good resilience to noise across the coupling regimes. The noise profile of the quantum measurements generating the GEE matrix elements slightly varies for each Pauli string but is independent of effective interaction strength. However, at considerably large coupling strength, the computational errors start to dominate over the theoretical ones, which is attributed to the increase of the GEE matrix rank with configuration complexity. The ZNE error mitigation method was applied to reduce the systematic error in the latter regimes and showed its effectiveness. However, the sampling noise remains significant for certain strong coupling regimes where the algorithm is sensitive to the degeneracy of the VQE ground state.

The qEOM approach admits further improvements and optimizations without compromising its advantages and major qualities. The most immediate developments are introducing self-consistency between the ground-state wave function and the excitation operator and transitioning from $npnh$ ans\"atze to the particle-vibration coupling ones, accentuating the effects of emergent collectivity. 
With the demonstrated quantum advantage, reasonable scaling with the system size, and noise resilience, the qEOM is one of the most promising methods to be implemented for realistic strongly interacting systems. Of particular interest are nuclear systems, which will be targeted in future work.


\begin{acknowledgments}

We are grateful for illuminating discussions with Denis Lacroix and Kyle Wendt and 
 funding from US-NSF through Grant PHY-2209376 and CAREER Grant PHY-1654379. E.L. acknowledges support from the GANIL Visitor Program. 
We appreciate cloud access to IBM quantum computers to run simulations for this work. 
The views expressed are those of the authors and do not reflect IBM's official policy or position.

\end{acknowledgments}
\bibliography{BibliographyMar2023}

\end{document}